\documentclass[pra,aps,showpacs,groupedaddress,superscriptaddress,twocolumn,toc=flat]{revtex4-1}

\usepackage[colorlinks=true,linkcolor=blue,urlcolor=blue,citecolor=blue]{hyperref}

\usepackage[utf8x]{inputenc}
\usepackage{color}
\usepackage{bbm} 

\usepackage{amsfonts,amsmath,amssymb,stmaryrd}

\usepackage{graphicx}
\usepackage{subfigure}  
\usepackage{bbm} 
\usepackage{hyperref}
\usepackage{epsfig}
\usepackage{mathrsfs}
\usepackage{verbatim}
\usepackage{centernot}
\usepackage{ulem}
\usepackage{nicefrac}

\renewcommand{\l}{\left(}
\renewcommand{\r}{\right)}

\newcommand{\bra}[1]{\langle#1|}
\newcommand{\ket}[1]{|#1\rangle}

\renewcommand{\ij}{{\langle \vec{i}, \vec{j} \rangle}}

\renewcommand{\H}{\hat{\mathcal{H}}}

\renewcommand{\c}{\hat{c}}
\renewcommand{\a}{\hat{a}}
\newcommand{\s}{\hat{s}}
\newcommand{\cd}{\hat{c}^\dagger}

\newcommand{\ad}{\hat{a}^\dagger}
\newcommand{\sd}{\hat{s}^\dagger}
\newcommand{\bd}{\hat{b}^\dagger}

\renewcommand{\b}{\hat{b}}
\newcommand{\hd}{\hat{h}^\dagger}
\newcommand{\h}{\hat{h}}

\newcommand{\hc}{\text{h.c.}}

\usepackage{array}

\usepackage{cancel,ifthen}
\newcommand{\cmnt}[2][NoInPuT]{\ifthenelse{\equal{#1}{NoInPuT}}{}{{\color{red}\sout{#1}}} {\color{blue} #2}}

\usepackage{bm}	
\renewcommand{\vec}[1]{\bm{#1}}

\begin{document}
\normalem	

\title{Rotational Resonances and Regge Trajectories in Lightly Doped Antiferromagnets}

\author{A. Bohrdt}
\email[Corresponding author email: ]{annabelle.bohrdt@tum.de}
\affiliation{Department of Physics and Institute for Advanced Study, Technical University of Munich, 85748 Garching, Germany}
\affiliation{Munich Center for Quantum Science and Technology (MCQST), Schellingstr. 4, D-80799 M\"unchen, Germany}
\address{ITAMP, Harvard-Smithsonian Center for Astrophysics, Cambridge, MA 02138, USA}
\affiliation{Department of Physics, Harvard University, Cambridge, Massachusetts 02138, USA}

\author{E. Demler}
\affiliation{Department of Physics, Harvard University, Cambridge, Massachusetts 02138, USA}

\author{F. Grusdt}
\affiliation{Department of Physics and Arnold Sommerfeld Center for Theoretical Physics (ASC), Ludwig-Maximilians-Universit\"at M\"unchen, Theresienstr. 37, M\"unchen D-80333, Germany}
\affiliation{Munich Center for Quantum Science and Technology (MCQST), Schellingstr. 4, D-80799 M\"unchen, Germany}


\date{\today}

\begin{abstract}
Understanding the nature of charge carriers in doped Mott insulators holds the key to unravelling puzzling properties of strongly correlated electron systems, including cuprate superconductors. Several theoretical models suggested that dopants can be understood as bound states of partons, the analogues of quarks in high-energy physics. However, direct signatures of spinon-chargon bound states are lacking, both in experiment and theory. Here we numerically identify long-lived rotational resonances at low doping, which directly reveal the microscopic structure of spinon-chargon bound states. Similar to Regge trajectories reflecting the quark structure of mesons, we establish a linear dependence of the rotational energy on the super-exchange coupling. Rotational excitations are strongly suppressed in standard angle-resolved photo-emission (ARPES) spectra, but we propose a multi-photon rotational extension of ARPES where they have strong spectral weight. Our findings suggest that multi-photon spectroscopy experiments should provide new insights into emergent universal features of strongly correlated electron systems.
\end{abstract}

\maketitle

\section{Introduction}
Our understanding of strongly correlated quantum matter often involves new emergent structures. For example, emergent gauge fields play a central role for understanding quantum spin liquids \cite{Wen2004}, and spin-charge separation in the one-dimensional Hubbard model can be related to the fractionalization of fermions into deconfined spinons and chargons \cite{Giamarchi2003,Kim1996,Vijayan2020}. The fate of those partons in dimensions higher than one remains unresolved. Theoretically and experimentally, one faces similar problems as in high-energy physics: the mathematical models are too challenging to solve and signatures of parton formation are often indirect or buried in complex observables. In this article we draw analogies to high energy physics and report on unambiguous signatures for parton structures in the two-dimensional (2D) $t-J$ model.

In quantum chromodynamics it is well established that directly observable nucleons are not the most elementary constituents of matter. The quark model introduced more than fifty years ago explains the larger class of mesons and baryons as composite objects consisting of two or three valence quarks. A smoking gun demonstration of the quark model was its ability to explain many additional resonances observed in collider experiments as ro-vibrational excitations of the fundamental parton configurations. In the quark model, many heavy mesons are thus understood as excited states of the fundamental mesons: they contain the same quark content but realize a higher vibrational state or have non-zero orbital angular momentum \cite{Micu1969,Olive2014}. 

A hallmark signature of rotational mesons comes from analysis of their mass, which can be related to the excitation energy of the pair of quarks relative to the lowest irrotational energy state. In a simplistic model, a meson can be described as a rigid line-like object with constant energy density per unit length, also known as ``string tension''. The two nearly massless quarks are located at the respective ends of this line and carry the (flavor) quantum numbers of the system. The relativistic expression for the energy of a rotating meson of this type scales linearly with the string tension and with the square root of its angular momentum \cite{Greensite2003}. The latter relation, known as Regge trajectory, can be directly probed in collider experiments and has been observed experimentally \cite{Bali2001}. It provides a strong indication that the observed mesons are bound states of partons. 

\begin{figure}
\centering
\epsfig{file=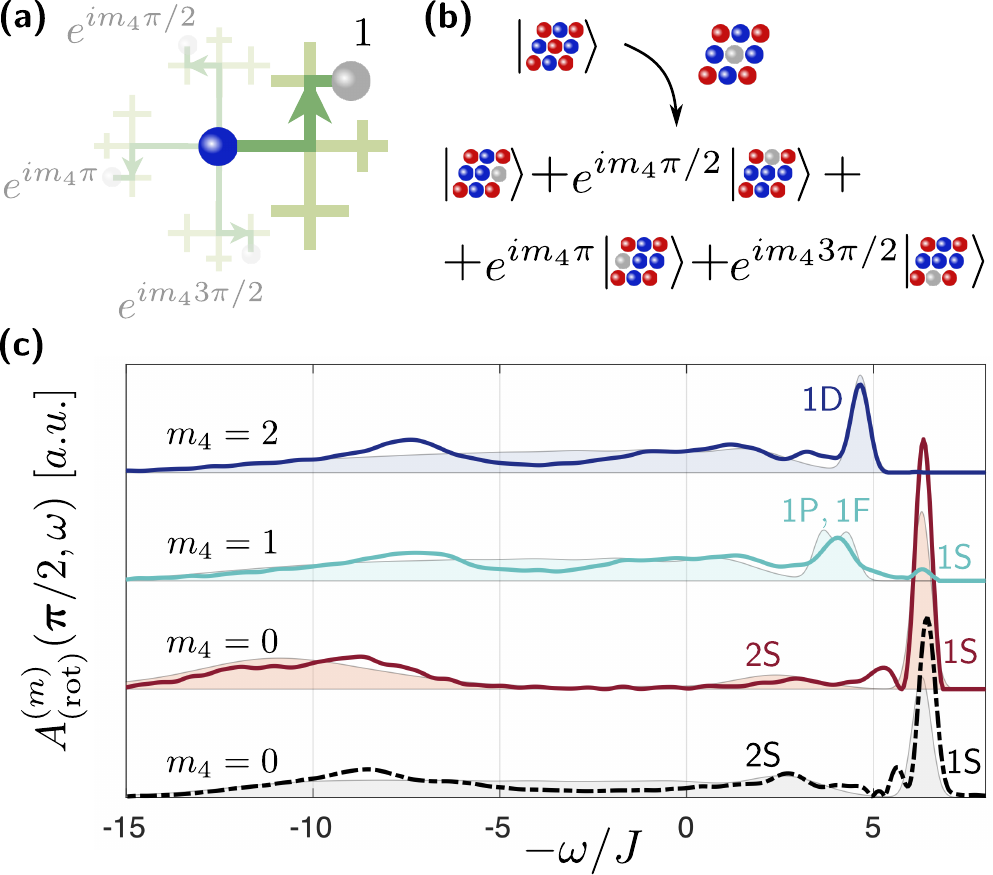, width=0.48\textwidth}
\caption{\textbf{Rotational meson resonances.} Bound states of spinons and chargons in a $C_4$-symmetric doped 2D AFM Mott insulators feature characteristic ro-vibrational excitations. (a) In an effective microscopic theory, the string with the light chargon (gray) can rotate around the heavy spinon (blue). (b) To detect rotational resonances, we propose a multi-photon ARPES scheme. Following the creation of a hole by a first photon, a second photon couples to lattice vibrations and excites rotational modes with $C_4$ angular momentum $m_4=0,1,2,3$. 
(c) Energy distribution curves for rotational ARPES spectra at the nodal point $\vec{k}= \vec{\pi}/2$ with different angular momenta, from top to bottom: $m_4=2,1,0$. The lowest (dash-dotted) curve corresponds to the usual ARPES spectrum with $m_4=0$. All spectra are normalized by their total area, $\int d\omega~ A_{{\rm (rot)}}^{(m)}$. The lowest mesonic resonances (ground state $\mathsf{1S}$, vibrational $\mathsf{2S}$ and rotational $\mathsf{1P,1D,1F}$ excited states) correspond to long-lived excited states. We performed time-dependent DMRG simulations for a $t-J$ model on a $4 \times 40$ cylinder, with $t=3 J$. The shaded areas correspond to toy model calculations (see methods \ref{MethodsC}) where we introduced small energy shifts and broadening as fit parameters. 
}
\label{figRotSetup}
\end{figure}

An idea almost as old as the problem of high-$T_c$ superconductivity itself comprises that strongly correlated electrons in these systems may be ruled by similar principles as high-energy physics \cite{Lee2006}. In analogy with quark confinement, B\'eran et al. suggested a description of a hole doped into a 2D antiferromagnet (AFM) in terms of a composite quasiparticle, consisting of two partons -- a chargon, carrying the charge quantum number and a spinon, carrying the spin quantum number -- bound together by ``an interaction obeying a string law'' \cite{Beran1996}. However, finding direct experimental or theoretical signatures for such structure has proven to be difficult. In angle resolved photo-emission spectra (ARPES) no sign of rotational resonances has been seen, and the nature of a possible first vibrational excitation is debated \cite{Dagotto1990,Leung1995,Mishchenko2001,Manousakis2007,Bohrdt2020}. 

Discerning the nature of charge carriers in lightly doped Mott insulators should provide a major boost to understanding properties of the underdoped cuprates and elucidating the origin of the pseudogap (PG) phase. In particular, it should provide a basis for constructing a consistent description of transport \cite{Badoux2016} and spectroscopy \cite{Shen2005} experiments. Several theoretical proposals involve emergent structures of partons, starting on a single dopant level \cite{Bulaevskii1968,Beran1996,Trugman1988,Manousakis2007}, to effective theories of the PG phase involving exotic composite Fermi liquids \cite{Anderson2007,Baskaran2007} and including fractionalized Fermi liquids \cite{Senthil2003} where deconfined spinons and chargons form electron-like bound states \cite{Punk2015PNASS,Sachdev2016,Zhang2020}. These scenarios may also explain the sudden and pronounced change of ARPES spectra \cite{Chen2019} and the carrier density \cite{Badoux2016} observed in cuprates around $p^* = 19 \%$ doping, as being related to an unbinding transition of spinons and chargons.

Here we provide strong numerical evidence that charge carriers in a lightly doped 2D AFM Mott insulators are comprised of partons, which are bound to each other, and exhibit telltale rotational excitations following Regge-like trajectories. We show that these rotational excitations are strongly suppressed in standard ARPES measurements and propose a multi-photon extension of ARPES imparting $C_4$-angular momentum into the system and allowing to access rotational excitations experimentally in solids \cite{Damascelli2003} or using ultracold atoms \cite{Bohrdt2018,Brown2019}. Our numerical DMRG simulations of the rotational ARPES spectra in the $t-J$ model, see Fig.~\ref{figRotSetup}, reveal narrow quasiparticle peaks at low excitation energies, which we interpret as a striking proof of the parton picture. Moreover, we describe the rotational resonances by a microscopic spinon-chargon toy model which explains the observed features without any free fit parameters.

\section{Rotational ARPES spectrum}
In traditional ARPES the spectral function $A(\vec{k},\omega) = - \pi^{-1} {\rm Im} \mathcal{G}(\vec{k},\omega)$ is measured which reveals information about the one-hole Green's function $\mathcal{G}(\vec{j},t) = \theta(t) \sum_\sigma \bra{\Psi_0} \cd_{\vec{j},\sigma}(t) \c_{\vec{0},\sigma}(0) \ket{\Psi_0}$; the latter describes how a fermion $\c_{\vec{j},\sigma}$ with spin $\sigma$ is removed from the initial state $\ket{\Psi_0}$ and leads to a hole propagating through the system. In the 2D Fermi-Hubbard model, believed to describe lightly doped copper oxides \cite{Lee2006}, a long-lived quasiparticle peak is found in $A(\vec{k},\omega)$ \cite{Wells1995,Ronning2005,Graf2007} which describes how a hole interacting with magnetic fluctuations forms a spin- or magnetic polaron \cite{Kane1989,Sachdev1989,Dagotto1990,Martinez1991,Liu1992,Koepsell2019} and moves through the surrounding AFM.

Our goal is to search for long-lived rotational excitations in the one-hole spectrum, which provide a direct route to reveal the composite nature of charge carriers in the Hubbard model. To couple to rotationally excited states one must impart discrete $C_4$ angular momentum into the system. However, the Green's function $\mathcal{G}(\vec{k},\omega)$ respects the symmetries of the underlying Hamiltonian: In this case we are particularly interested in the discrete rotational $C_4$ symmetry of the Hubbard model, which is unbroken in the undoped parent AFM $\ket{\Psi_0}$. Hence, for $C_4$ invariant momenta (C4IM) in the magnetic Brillouin zone (MBZ) no angular momentum transfer is allowed and rotational excitations have no weight in the traditional ARPES spectrum $A(\vec{k},\omega)$. 

For non-C4IM, lattice effects can in principle impart $C_4$ angular momentum into the system. However since the Green's function couples to the center-of-mass momentum $\vec{k}$ of the hole, the spectral weight of rotational states is still expected to be strongly suppressed if the effective masses of the two supposed partons are significantly different. In this limit, the lighter parton rotates around the heavier parton which carries most of the linear momentum $\vec{k}$, thus suppressing couplings of $\vec{k}$ to the \emph{relative} angular momentum of the two partons. We confirm this intuition for an analytically solvable toy-model in one dimension (see supplements \ref{SuppMat1Dmes}). The Hubbard model in cuprates, with super-exchange coupling $J \approx t / 3$, is in such a regime where significantly different parton masses $\simeq 1/J$ and $\simeq 1/t$ are expected. 

To allow significant overlap with possible rotational excitations, we devise a rotational extension of ARPES where angular momentum is directly imparted into the system, even at C4IM. The simplest term creating an excitation with discrete angular momentum $m_4=0,1,2,3$, spin $\sigma$, charge one, and total momentum $\vec{k}$ is given by
\begin{equation}
\hat{R}_{m_4,\sigma}(\vec{k}) =  \sum_{\vec{j}} \frac{e^{- i \vec{k} \cdot \vec{j}}}{\sqrt{V}}  \sum_{\vec{i}: \ij} e^{i m_4 \varphi_{\vec{i} - \vec{j}}}  \sum_{\sigma'} \cd_{\vec{j},\sigma'} \c_{\vec{i},\sigma'} \c_{\vec{j},\sigma},
\label{eqDefRmk}
\end{equation}
with $\varphi_{\vec{r}} = {\rm arg}(\vec{r})$ the polar angle of $\vec{r}$. The action of this operator on a product N\'eel state, $\hat{R}_{m_4,\sigma} \ket{{\rm N}}$, is illustrated in Fig.~\ref{figRotSetup} (b): Here the second and third fermion operators in $\hat{R}_{m_4,\sigma}$ create a string-like excitation with $C_4$ angular momentum $m_4$ and non-zero overlap to the rotational states predicted for a hole in an Ising AFM \cite{Grusdt2018PRX}.

Instead of the usual Green's function, we consider the rotational Green's function
\begin{equation}
\mathcal{G}_{\rm rot}^{(m_4)}(\vec{k},t) = \theta(t) \sum_\sigma \bra{\Psi_0} \hat{R}^\dagger_{m_4,\sigma}(\vec{k},t) \hat{R}_{m_4,\sigma}(\vec{k},0) \ket{\Psi_0},
\end{equation}
which we calculate by time-dependent DMRG (see methods \ref{MethodsB}) \cite{Paeckel2019a,Kjall2013,Zaletel2015}. The corresponding rotational spectrum, $- \pi^{-1} {\rm Im} \mathcal{G}_{\rm rot}^{(m_4)}(\vec{k},\omega)$, in Lehmann representation is
\begin{equation}
 A_{\rm rot}^{(m_4)}(\vec{k},\omega) =  \sum_{\sigma, n>0} \delta \l \omega-E_n+E_0 \r  | \bra{\Psi_n} \hat{R}_{m_4,\sigma}(\vec{k})  \ket{\Psi_0} |^2,
 \label{eqDefArot}
\end{equation}
where $\ket{\Psi_0}$ ($E_0$) is the correlated ground state (energy) and $\ket{\Psi_n}$ ($E_n$) for $n>0$ are the eigenstates (eigenenergies) with an added hole. Hence, if long-lived rotational excitations exist, they manifest in pronounced quasiparticle peaks in the rotational ARPES spectrum in Eq.~\eqref{eqDefArot}. For $m_4=0$ the same selection rules apply as for the conventional ARPES spectrum and the same states contribute, but with modified spectral weights. 

The rotational spectrum can be experimentally measured using a multi-photon extension of ARPES. We propose to use one set of beams for lattice modulation, which imparts angular momentum into the system by coupling to specific phonon modes in solids \cite{Devereaux1994} or directly by modulating the optical potential with appropriate phases in ultracold atoms \cite{Bloch2008}. The other beam is the usual ARPES beam which creates the hole excitation. Details of our scheme are provided in the methods \ref{MethodsA}.

\section{Rotational resonances}
Now we present our numerical results obtained for one hole doped into a 2D AFM Mott insulator. Specifically, we considered the $t-J$ model on extended four-leg cylinders and for $t/J=3$, the experimentally most relevant value for cuprates (see methods \ref{MethodsB}). In Fig.~\ref{figRotSetup} (c) we show numerically obtained spectra (energy distribution curves) at the nodal point $\vec{k}= \vec{\pi}/2$, with $\vec{\pi} = (\pi,\pi)$. For $m_4=0$ (red line, second from bottom) the rotational spectrum shows the same quasiparticle peak as the conventional ARPES spectrum (black, bottom line), at the same energy. This peak, labeled $\mathsf{1S}$, corresponds to the magnetic polaron ground state \cite{Bohrdt2020}. A possible excited state is also visible at $m_4=0$, which we label $\mathsf{2S}$ and which has previously been argued to correspond to the first vibrational excitation of the magnetic polaron \cite{Dagotto1990,Leung1995,Mishchenko2001,Manousakis2007,Bohrdt2020}. The $\mathsf{2S}$ state has a reduced spectral weight in the rotational ARPES spectrum, where it is difficult to identify at all. 

Much clearer indications for long-lived excitations of magnetic polarons can be found in the non-trivial rotational ARPES spectra with $m_4 \neq 0$. For $m_4=2$ (top, blue curve in Fig.~\ref{figRotSetup}) we find a pronounced quasiparticle peak corresponding to an excitation energy $\Delta E \sim 1.7 J$. Remarkably, no significant spectral weight appears below this energy, in particular we find zero spectral weight at the polaron ground state energy. We note that this is not simply a consequence of selection rules: Firstly, the nodal point does not correspond to a C4IM, not even in the reduced MBZ. Secondly, the AFM has gapless magnon modes which should in principle allow to carry away angular momentum and allow an excited magnetic polaron to decay to its ground state. Based on these observations, we identify the resonance found at $m_4=2$ with a $\mathsf{1D}$ excited state of the magnetic polaron. 

Similarly, the rotational spectrum with $m_4=1$ features a pronounced peak at a slightly higher excitation energy $\Delta E \sim 2.3 J$ above the ground state (second from top, turquoise curve in Fig.~\ref{figRotSetup}). In this case we find weak hybridization with the $\mathsf{1S}$ state, as indicated by a small quasiparticle peak at the ground state energy. Based on its quantum numbers, we identify the new excited state as $\mathsf{1P}$. By applying a combination of time-reversal and inversion symmetry, it follows that the $m_4=1$ and $m_4=3$ rotational spectra coincide exactly. Hence the $\mathsf{1P}$ state is associated with a degenerate $\mathsf{1F}$ state at $m_4=3$. 

Our observation of long-lived quasiparticle peaks in the rotational spectrum provides a direct indication that mobile holes in lightly doped AFM Mott insulators have a discrete internal structure. To understand our reasoning, consider a theoretical model of magnetic polarons without a rigid internal structure. In this case, the action of the operator $\hat{R}_{m_4,\sigma}(\vec{k})$ in the rotational Green's function would generically be expected to have two separate effects: The first fermion operator $\c_{\vec{j},\sigma}$ in Eq.~\eqref{eqDefRmk} creates a mobile hole with a large overlap to the structureless magnetic polaron. The subsequent pair of fermion operators, $\sum_{\sigma'} \cd_{\vec{j},\sigma'} \c_{\vec{i},\sigma'}$ in Eq.~\eqref{eqDefRmk}, then couples to the surrounding spins and creates separate magnon excitations. In this case one would expect $A_{\rm rot}^{(m_4)}(\vec{k},\omega)$ to become a convolution of a polaron and a magnon contribution, possibly renormalized weakly by interaction effects. This would lead to a broad and mostly featureless spectrum -- in stark contrast with our numerical findings in Fig.~\ref{figRotSetup} (c).

\section{Regge-like trajectories}
B\'eran et al. \cite{Beran1996} have suggested that mobile holes in an AFM Mott insulator can be described as mesonic bound states of two strongly interacting partons, a light chargon and a heavy spinon, see also \cite{Laughlin1997,Grusdt2018PRX}. In that case the operators $\hat{R}_{m_4,\sigma}(\vec{k})$ should create rotational excitations, which explains the peaks we found in the rotational spectra in Fig.~\ref{figRotSetup}. Now we study how the excitation energies $\Delta E$ of these rotational peaks, as well as the first vibrational peak, depend on the underlying coupling strength $J/t$ in the system, which gives further insights into the nature of the bound state. 

\begin{figure}[t!]
\centering
\epsfig{file=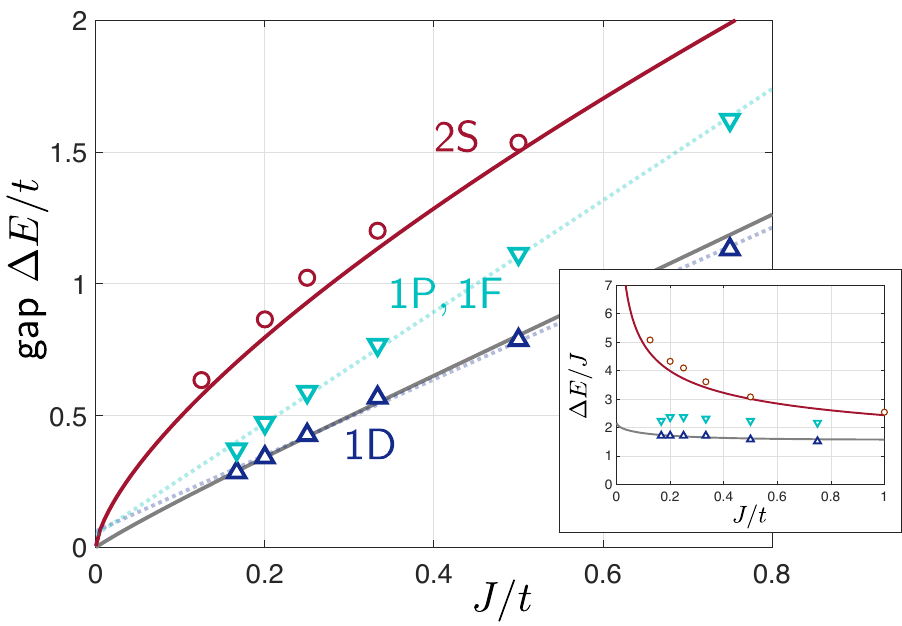, width=0.49\textwidth}
\caption{\textbf{Meson Regge trajectories.} We show the dependence of the excitation gap $\Delta E$ at the nodal point $\vec{k}=\vec{\pi}/2$ on the super-exchange energy $J$. The gap was extracted from peak positions in numerically obtained spectra. The low-lying rotational resonances ($\mathsf{1P, 1D, 1F}$) have a gap scaling linearly with $J$ (light dotted lines denote linear fits: $\Delta E_{\mathsf{1D}} = 1.44 J + 0.061 t$ and $\Delta E_{\mathsf{1P,1F}} = 2.12 J + 0.047 t$). The gap to the first vibrational peak ($\mathsf{2S}$) scales with $t^{1/3} J^{2/3}$ \cite{Bohrdt2020}. Solid lines are parameter-free calculations neglecting spinon dynamics, see methods \ref{MethodsC}. The inset shows the same data, but with energy measured in units of $J$ instead of $t$.
}
\label{figReggeTraj}
\end{figure}

In Fig.~\ref{figReggeTraj} we numerically extracted the positions of the peaks from frequency cuts $A_{\rm rot}^{(m_4)}(\vec{\pi}/2,\omega)$ of rotational spectra at the nodal point, for different values of $J/t$. We find that the positions of the rotational peaks scale linearly with the spin exchange 
\begin{equation}
\Delta E_{\rm rot} \simeq J,
\label{eqErotScaling}
\end{equation}
whereas the gap to the vibrational excitation $\mathsf{2S}$ has a characteristic power-law dependence on $t$ and $J$ \cite{Bohrdt2020}, 
\begin{equation}
\Delta E_{\rm vib} \simeq t^{1/3} J^{2/3}.
\label{eqEvibScaling}
\end{equation} 

These scaling behaviors can be explained by a simplistic meson model~\cite{Grusdt2018PRX}: In this model the two partons are connected by a line-like object on the square lattice with constant energy density $dE/d\ell$. This string tension must be proportional to the spin exchange energy $dE/d\ell \propto J$ to obtain the observed scaling laws in Eqs.~\eqref{eqErotScaling}, \eqref{eqEvibScaling}.

Since $J$ corresponds to the string tension between the two partons, Eq.~\eqref{eqErotScaling} resembles the celebrated Regge formula from particle physics, which relates the meson excitation energy to  its angular momentum and the underlying string tension \cite{Greensite2003}. While high-energy experiments cannot tune the string tension, which is determined by the coupling constant $g$ of quantum chromodynamics, cold atom quantum simulators \cite{Cheuk2016,Mazurenko2017,Koepsell2019,Brown2019} can be used to tune the coupling $J/t$ in the Hubbard model and directly measure the Regge-like trajectories we predict for rotational excitations in Fig.~\ref{figReggeTraj}.

In further analogy with the Regge formula from high-energy physics, we can study the dependence of the meson excitation energy $\Delta E$ on its angular momentum $m_4$. While quarks in high-energy physics are described in a continuous space-time, lattice effects are strong in the Hubbard or $t-J$ models we consider. As a result, the simplistic meson model from Ref.~\cite{Grusdt2018PRX} predicts that all rotational states with $m_4 \neq 0$ should be degenerate when $J/t \ll 1$, and be located between purely vibrational states in this limit. Refined meson models predict small splittings of the rotational lines, however. We confirm in Fig.~\ref{figReggeTraj} that all rotational excitations are close in energy, and well separated from the first vibrational peak.

\begin{figure}[t!]
\centering
\epsfig{file=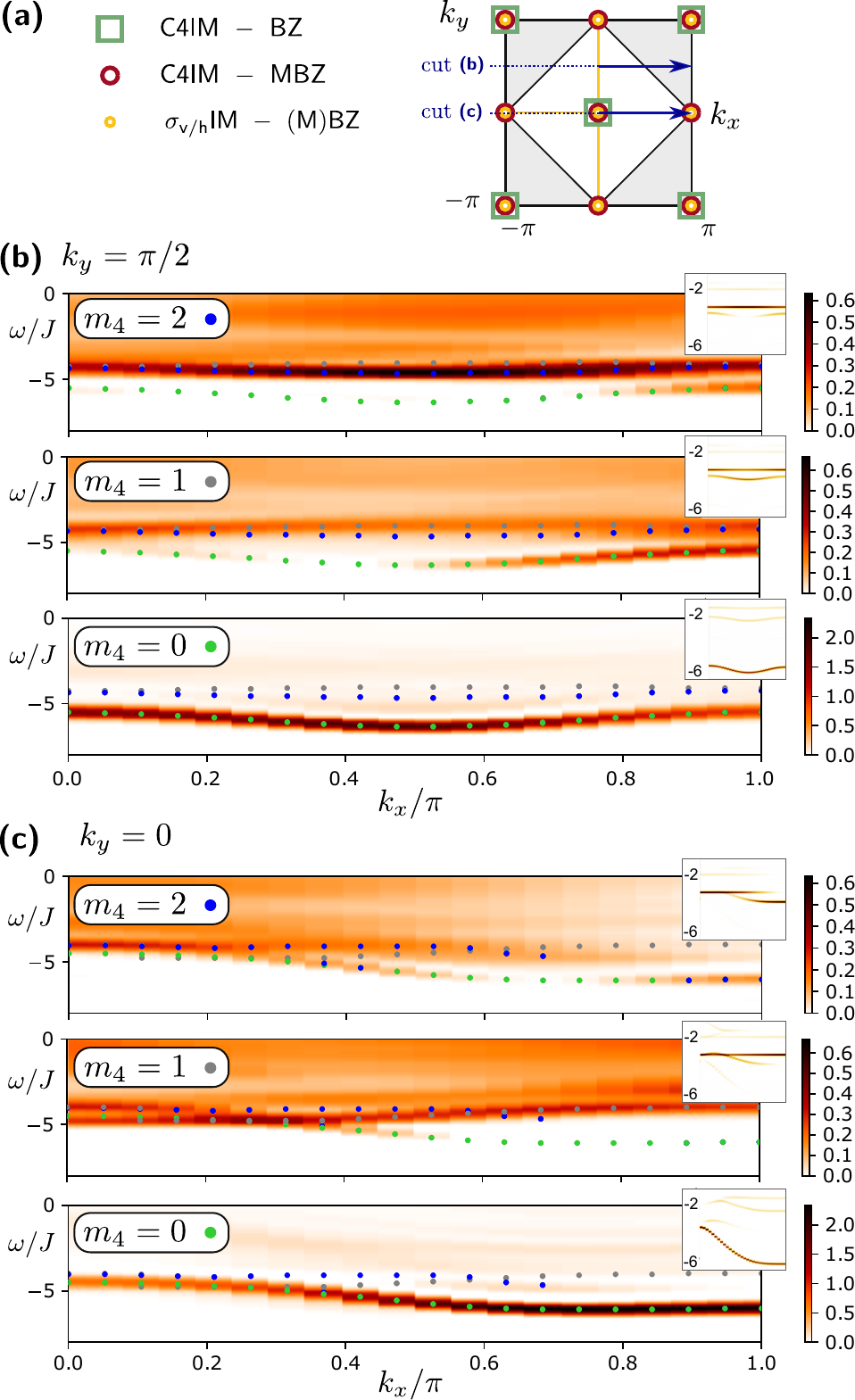, width=0.49\textwidth}
\caption{\textbf{Rotational spectra and meson dispersion.} We show full rotational ARPES spectra $A_{\rm rot}^{(m_4)}(\omega,\vec{k})$ for momenta $\vec{k}$ along the cuts in the Brillouin zone indicated by blue arrows in (a). For a given value of $m_4=0$ ($m_4=1,2$), the fitted peak positions of the low-lying resonances are indicated by green (gray, blue) dots in (b) and (c). The numerics were carried out on elongated $L_x=40 \times 4$-leg cylinders, for a $t-J$ model with $t/J=3$. At high-symmetry momenta in the Brillouin zone, see (a), selection rules explain why some meson resonances are invisible. The insets show the same spectral cuts in the spinon-chargon toy model, where discrete delta-functions were replaced by weakly broadened Gaussian peaks but without any further fitting parameters.
}
\label{figSpectralLines}
\end{figure}

\section{Meson dispersion}
The dependence of the peak position on the momentum $\vec{k}$ enables further insights into the properties of the mesonic bound states describing mobile holes in lightly doped AFM Mott insulators. In Fig.~\ref{figSpectralLines} we show momentum cuts of the rotational ARPES spectrum, along lines indicated in part (a), for a fixed ratio of $t/J=3$. 

In order to simplify comparison between spectra at different angular momenta, we indicate the numerically extracted dispersions of all mesonic resonances by dots. Different dot colors denote the angular momentum $m_4$ where the respective peak is most pronounced. Since $m_4$ is only a conserved quantum number at C4IM, at generic momenta $\vec{k}$ most mesonic resonances are visible for different values of $m_4$ at the same time. In addition, even at C4IM our numerics indicate further weak hybridization of rotational states which is caused by the broken $C_4$ rotational symmetry in the elongated four-leg cylinders we consider. For example, at $\vec{k}=(\pi,0)$ the $\mathsf{1S}$ state has non-zero spectral weight for $m_4=2$. On the other hand, the strict vertical and horizontal reflection symmetries $\sigma_{\rm v/h}$ of our finite-size cylinders prevent hybridization of states with even and odd $m_4$, respectively. 

Our main findings from the meson dispersions in Fig.~\ref{figSpectralLines} are as follows. For almost all momenta, the lowest peak corresponds to angular momentum $m_4=0$ (green dots). This peak shows the strongest dispersion, with a minimum at the nodal point. In contrast, the rotational meson resonances show significantly reduced dispersion and the locations of dispersion minima depend strongly on the value of $m_4$. These features are explained by the spinon-chargon toy model presented in the methods section \ref{MethodsC}.

\section{Measurement scheme}
We expect that the rotational ARPES spectra we predict can be experimentally measured in cuprate compounds. Specifically, we propose to perform multi-photon spectroscopy which combines the usual ARPES beam with a periodic lattice modulation. To realize the latter, we suggest to drive the buckling phonon modes in copper-oxide layers, in particular the Raman active $A_{\rm 1g}$ and $B_{\rm 1g}$ modes in YBCO \cite{Devereaux1994,Rosch2004}. Overall this leads to an effective three-photon transition, consisting of a pair of Raman lasers and the usual ARPES beam. As we show in the methods \ref{MethodsA}, the $s$- and $d$-wave symmetries of the $A_{\rm 1g}$ and $B_{\rm 1g}$ modes, respectively, allow to measure $m_4=0$ and $m_4=2$ rotational ARPES spectra. 

We also show in the methods \ref{MethodsA} that a similar rotational extension of scanning-tunneling microscopy (STM) can be envisioned. Even though rotational STM spectra lack momentum resolution, we demonstrate that the predicted weakly dispersive long-lived rotational meson resonances in Fig.~\ref{figSpectralLines} can be resolved. 

Finally, we expect that analogous measurements will become possible in ultracold atoms in the immediate future \cite{Brown2019,Kollath2007,Bohrdt2018}. There, the required angular momentum can be imparted into the system directly through shaking of the optical lattice and without phonons, resulting in a two-photon spectroscopy scheme; see methods \ref{MethodsA} for details.

\section{Discussion and Outlook}
We have proposed a rotational extension of ARPES, which we used to predict the previously unknown long-lived rotational excitations of charge carriers in lightly doped 2D AFM Mott insulators. Our finding of pronounced quasiparticle peaks in the rotational spectra allow to conclude that strong interactions between spin and charge must be present. By analyzing Regge-like trajectories, describing the dependence of the excitation energy on the super-exchange $J$, we found compelling evidence that mobile holes in lightly doped AFM Mott insulators have a rich internal structure and can be understood as spinon-chargon bound states. This finding is further supported by the good agreement we report with a microscopic toy model of spinons and chargons connected by a string on the square lattice. 

We will extend our numerical analysis, obtained for the one-hole doped $t-J$ model so far, to study the closely related 2D Hubbard model and analyze effects of weak non-zero doping next. We do not expect the clear numerical signatures obtained so far -- which we also confirm in exact diagonalization of $4 \times 4$ systems in the supplements \ref{SuppMatAddNum} -- to change significantly in this case.  

Our research provides the most direct evidence yet for the decades old idea \cite{Beran1996,Laughlin1997} that the physics of lightly doped 2D AFM Mott insulators -- and by extension high-temperature superconductors -- is captured by emergent partons. In particular our results support the picture of the pseudogap phase in cuprates as a liquid of fermionic mesons, modeling charge carriers as bound states of spinons and chargons. We emphasize the importance of a direct experimental confirmation that charge carriers have a rich internal structure: An observation of the long-lived rotational resonances we predict would provide a strong indication that this picture is correct. 

The meson structure of charge carriers in lightly doped 2D AFM Mott insulators may have further theoretical implications. On the one hand it may contribute to our understanding how stripes form at low temperatures \cite{Grusdt2020}, or shed new light on the pairing mechanism underlying high-temperature superconductivity in cuprates. On the other hand, understanding possible un-binding transitions of spinons and chargons may contribute to a deeper understanding of the rich phase diagram of cuprates. An interesting future direction would be to explore how the parton picture relates to the sudden change of carrier properties observed around a critical doping $p^*\approx 19\%$ \cite{Tallon2001,Badoux2016,Chen2019}, see also \cite{Koepsell2020a}.

\newpage

\section{Methods}

\subsection{Multi-photon lattice modulation spectroscopy}
\label{MethodsA}
To detect rotational excitations experimentally, we propose a general lattice modulation scheme. The lattice modulation causes a two-photon side-band (or, depending on the coupling scheme, a Raman side-band) in the ARPES spectrum, shifted by the rotational excitation energy. We discuss possible implementations of the rotational ARPES scheme for ultracold atoms in quantum gas microscopes \cite{Gross2017}, and in solids where three photon beams are required to Raman-couple to suitable phonon modes. Finally we show how scanning-tunneling microscopy (STM) can be combined with the same lattice shaking scheme to probe the rotational states without momentum resolution.

\subsubsection{General scheme}
To describe the lattice modulation scheme, we denote by $\H_{0}$ the underlying system Hamiltonian ($t-J$ or Fermi-Hubbard model; additional terms with the same discrete lattice symmetries are also allowed). $\H_{0}$ determines the dynamics of the underlying particles $\c_{\vec{j},\sigma}$ in the system. By $\hat{V}_{\rm A}(t)$ we denote the applied perturbation whose linear response yields the ARPES signal: 
\begin{equation}
 \hat{V}_{\rm A}(t) = - \delta t_{\rm A} \sin( \omega_{\rm A} t)  \sum_{\vec{j},\sigma} \l  \ad_{\vec{j},\sigma} \c_{\vec{j},\sigma} + \hc \r.
 \label{eqDefVA}
\end{equation}
Here $\a_{\vec{j},\sigma}$ denotes the "photo electron" channel in the ARPES sequence. 

We assume that the detection system is governed by a non-interacting Hamiltonian,
\begin{equation}
 \H_a = \sum_{\vec{k},\sigma} \epsilon_{\vec{k},\sigma} \ad_{\vec{k},\sigma} \a_{\vec{k},\sigma},
\end{equation}
with a known dispersion $\epsilon_{\vec{k},\sigma}$. Further, we require that the energy, momentum and spin of the "photo electrons" created by $\ad_{\vec{k},\sigma}$ can be experimentally measured. The dispersion relation $\epsilon_{\vec{k},\sigma}$ is of the form 
\begin{equation}
 \epsilon_{\vec{k},\sigma} = \Delta + \delta \epsilon_{\vec{k},\sigma},
\end{equation}
where $\Delta$ is an overall energy offset and ${\rm min}_{\vec{k}} \delta \epsilon_{\vec{k},\sigma} = 0$. Note that $\Delta$ corresponds to the minimum energy required to remove a fermion from the system.
 
\begin{figure}[t!]
\centering
\epsfig{file=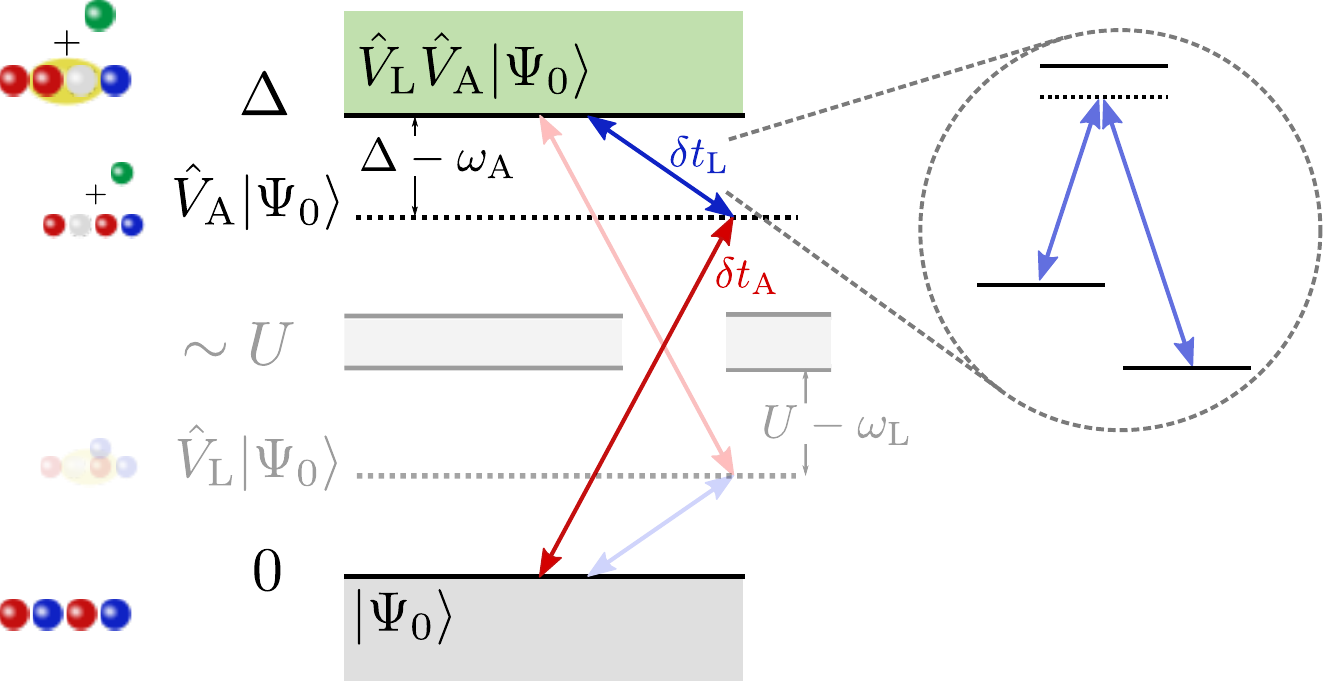, width=0.48\textwidth}
\caption{\textbf{Multi-photon rotational ARPES scheme.} To probe the rotational response of (doped) anti-ferromagnets, we propose to use a red-detuned ARPES beam $\delta t_{\rm A}$ and supplement the missing energy to extract a "photo electron" (green) by the lattice modulation $\delta t_{\rm L}$, i.e. $\omega_{\rm A} + \omega_{\rm L} \approx \Delta$. We assume that the lattice modulation frequency $\omega_{\rm L}$ is far-off resonant from the doublon-hole resonance at energy $\sim U$. In solids, the lattice modulation can be realized by Raman-coupling to suitable phonon modes, see inset.
}
\label{figRARPES}
\end{figure}

The second ingredient for our scheme is a super-lattice modulation, which periodically changes the tunnel coupling between neighboring sites $\vec{i}$ and $\vec{j}$:
\begin{equation}
 \hat{V}_{\rm L}(t) = - \sum_{\ij}  \delta t_{\rm L} \sin ( \omega_{\rm L} t + \phi_{\ij} )  \sum_\sigma \l \cd_{\vec{j},\sigma} \c_{\vec{i},\sigma} + \hc \r.
 \label{eqDefVL}
\end{equation}
We assume that the phases $\phi_\ij$ on different bonds $\ij$ can be controlled by the spatial structure of the lattice modulation. If resonant, the perturbation \eqref{eqDefVL} creates doublon-hole pairs with energy $\approx U \gg t,J$. 

The basic strategy to detect rotational excitations $m=0,1,2,3$ is to consider a two-photon transition involving both drives $\hat{V}_{\rm A}(t)$ and $\hat{V}_{\rm L}(t)$. The allowed transitions are shown in Fig.~\ref{figRARPES}. We assume that both one-photon transitions are off-resonant, i.e. 
\begin{flalign}
 |\delta t_{\rm A}| &\ll \underbrace{\Delta -  \omega_{\rm A}}_{> 0}, \\
  |\delta t_{\rm L}| &\ll | U - \omega_{\rm L} |.
\end{flalign}
In this case, "photo electrons" $\a_{\vec{k},\sigma}$ can only be created by a two-photon process. In the linear response regime, i.e. for sufficiently weak modulations $\delta t_{\rm A,L}$, the probability per time $\gamma_{\rm AL}$ for the two-photon transition is given by
\begin{multline}
\gamma_{\rm AL}(\vec{k}) = 2 \pi ~ \delta \bigl( \omega_{\rm L}+\omega_{\rm A} - ( E_f(\vec{k}) - E_i ) \bigr) \times \\
\biggl| \sum_{r} \biggl\{ \frac{\bra{\psi_f(\vec{k})} \hat{V}_{\rm A} \ket{r} \bra{r} \hat{V}_{\rm L} \ket{\psi_i}}{\omega_{\rm L} - (E_r - E_i) + i \Gamma_r}  + \frac{\bra{\psi_f(\vec{k})} \hat{V}_{\rm L} \ket{r} \bra{r} \hat{V}_{\rm A} \ket{\psi_i}}{\omega_{\rm A} - (E_r - E_i) + i \Gamma_r}  \biggr\}   \biggr|^2,
\label{eqGammaAL}
\end{multline}
where we defined for $\mu={\rm A, L}$ the time-independent perturbations $\hat{V}_\mu$ by the relation
\begin{equation}
 \hat{V}_\mu(t) \equiv e^{ - i \omega_\mu t} \hat{V}_\mu + \hc ~.
\end{equation}
In Eq.~\eqref{eqGammaAL}, $E_f(\vec{k})$ denotes the energy of the final state $\ket{\psi_f(\vec{k})} = \ket{\Psi_f}\ket{\vec{k},\sigma}$ with a "photo electron" $\a_{\vec{k},\sigma}$ with momentum $\vec{k}$, $E_i$ is the energy of the initial state $\ket{\psi_i} = \ket{\Psi_0} \ket{0}$ without "photo electrons"; $\ket{\Psi_f}$ and $\ket{\Psi_0}$ denote the final and ground states of the many-body system. The sum $\sum_r$ is taken over all possible intermediate states $\ket{r}$ with energy $E_r$ and inverse lifetime $\Gamma_r$ (where $\Gamma_r \to 0^+$ for infinite lifetimes); see e.g.~\cite{Long2002}.

The rotational spectra defined in the main text correspond to the response $\hat{V}_{\rm L} \hat{V}_{\rm A}$, where the ARPES transition takes place before the lattice modulation $\hat{V}_{\rm L}$ is applied. This ensures that $\hat{V}_{\rm L}$ transfers discrete orbital ($\hat{C}_4$) angular momentum to the many-body system but not to the "photo electrons" $\a_{\vec{k},\sigma}$; note that $[\hat{V}_{\rm L}, \hat{V}_{\rm A}] \neq 0$ in general. To ensure that the two-photon response in Eq.~\eqref{eqGammaAL} is dominated by terms where $\hat{V}_{\rm A}$ is applied first, we require that
\begin{equation}
 0 < \Delta - \omega_{\rm A}  \ll |U-\omega_{\rm L}|.
\end{equation}
I.e. the ARPES beam is close to resonance while the lattice modulation is far off resonant, see Fig.~\ref{figRARPES}. 

The response in Eq.~\eqref{eqGammaAL} simplifies further, if we assume that the intermediate states $\ket{r}$ are detuned sufficiently far, 
\begin{equation}
|\omega_{\rm A} - \Delta| \gg t, J,
\end{equation}
to neglect any dependence on the specifics of the intermediate states: $(E_r - E_i) \approx  \Delta$. This yields
\begin{multline}
\gamma_{\rm AL}(\vec{k}) = 2 \pi ~ \frac{| \bra{\psi_f(\vec{k})} \hat{V}_{\rm L} \hat{V}_{\rm A} \ket{\psi_i}|^2 }{ (\omega_{\rm A} - \Delta)^2 }  \times \\
   \delta \bigl( \omega_{\rm L}+\omega_{\rm A} - ( E_f(\vec{k}) - E_i ) \bigr).
   \label{eqALres}
\end{multline}
Note however, that this approximation does not affect the energy delta-function in Eq.~\eqref{eqGammaAL}.

Alternatively, if we assume that the quasiparticle lifetimes of the intermediate states are all relatively short,
\begin{equation}
 	\Gamma_r \approx \Gamma \geq |E_r - E_i| \approx \Delta,
	\label{eqGammaApprx}
\end{equation}
we can again sum over all intermediate states and get
\begin{multline}
\gamma_{\rm AL}(\vec{k}) 
\approx 2 \pi ~ \frac{| \bra{\psi_f(\vec{k})} \hat{V}_{\rm L} \hat{V}_{\rm A} \ket{\psi_i}|^2 }{ | \omega_{\rm A} - \Delta + i \Gamma |^2 }  \times \\
   \delta \bigl( \omega_{\rm L}+\omega_{\rm A} - ( E_f(\vec{k}) - E_i ) \bigr).
   \label{eqALresGamma}
\end{multline}

Finally, for $m=0,1,2,3$ we consider a specific lattice modulation $\hat{V}_{\rm L}$, with the following phases:
\begin{flalign}
 \phi_{\langle \vec{j}, \vec{j} + \vec{e}_x \rangle} &= \pi m  j_x, \label{eqPhaseMod1}\\
 \phi_{\langle \vec{j}, \vec{j} + \vec{e}_y \rangle} &= \frac{\pi}{2} m + \pi m j_y.
 \label{eqPhaseMod2}
\end{flalign}
For this choice one obtains
\begin{multline}
 \hat{V}_{\rm L} \hat{V}_{\rm A} \ket{\psi_i} = \frac{\delta t_{\rm L} \delta t_{\rm A}}{4} \sum_{\vec{k}, \sigma} \ket{\vec{k},\sigma} \frac{1}{\sqrt{V}} \sum_{\vec{j}} e^{- i (\vec{k} + \vec{\pi} m) \cdot \vec{j}}  \\
 \sum_{\vec{i}: \ij} e^{i m \varphi_{\vec{i} - \vec{j}}} \sum_{\sigma'} \cd_{\vec{j},\sigma'} \c_{\vec{i},\sigma'} \c_{\vec{j},\sigma} \ket{\Psi_0},
\end{multline}
where $V=L^2$ is the area of the system, $\vec{\pi}=(\pi,\pi)^T$ and $\varphi_{\vec{i} - \vec{j}}$ is the polar angle of the vector $\vec{i} - \vec{j}$. Plugging this result into Eq.~\eqref{eqALres} yields the rotational ARPES spectrum discussed in the main text in spectral, or Lehmann, representation:
\begin{multline}
\gamma_{\rm AL}^{(m)}(\vec{k}) \propto \sum_{\sigma=\uparrow, \downarrow} \sum_{\ket{f}} |\bra{f} \hat{R}_{m,\sigma}(\vec{k} + \vec{\pi} m) \ket{\Psi_0} |^2 \\
\times    \delta \bigl( \omega_{\rm L}+\omega_{\rm A} - ( E_f(\vec{k}) - E_i ) \bigr).
\end{multline}
Here $\hat{R}_{m,\sigma}(\vec{k})$ creates rotational states with total momentum $\vec{k}$, as defined in Eq.~\eqref{eqDefRmk} in the main text; we summed over all final states $\ket{f}$ of the many-body system contributing to instances where a "photo electron" with momentum $\vec{k}$ and arbitrary spin $\sigma$ is detected. 

The following corrections to the scheme presented above can be expected: (i) In a Hubbard model, the lattice modulation may lead to recombinations of virtual doublon-hole pairs. Such processes correspond to intermediate states $\ket{r}$ with energies $E_r - E_i \simeq J$, i.e. they are close to resonance. However, the spectral weights of such processes are strongly suppressed by the probability for virtual doublon-hole pairs which scales as $\simeq (U/t)^2 \ll 1$. (ii) With the tunneling $t$, the spin-exchange $J$ is modulated accordingly at the frequency $\omega_{\rm L}$. This leads to another contribution to the response $\gamma_{\rm AL}^{(m)}$. Since $t < J$ it follows that $\delta t < \delta J$ and we neglect this sub-dominant contribution. As can be seen from the delta-function in Eq.~\eqref{eqGammaAL}, all these corrections can only modify the respective spectral weights but not the positions or symmetry properties of the expected spectral lines.

\subsubsection{Realizations with ultracold atoms}
To realize the two-photon scheme introduced above for ultracold atoms in optical lattices, the "photo electron" channel must be replaced by non-interacting atomic states which are experimentally accessible. One option is to use a third internal atomic state, which may occupy the same lattice, but does not interact with the two spin states $\sigma=\uparrow, \downarrow$ used to realize the doped Hubbard model. Such settings have been proposed \cite{Dao2007} and realized in the continuum \cite{Stewart2008} and in lattice systems under a quantum gas microscope \cite{Brown2019}. A second option is to use spatially separated states to realize the ARPES channel \cite{Kollath2007,Bohrdt2018}: Specifically, a decoupled layer in the optical lattice, adjacent to the physical system, can be utilized in quantum gas microscopes \cite{Preiss2015,Koepsell2020PRL}. The offset $\Delta$ corresponds to the energy offset between the physics and detection layer, and can be independently tuned in ultracold atom systems. 

In all these systems, momentum resolved detection of the $\a_{\vec{k},\sigma}$ particles is required. In addition to the momentum resolution itself, this yields the energy of the emitted "photo electron" provided its dispersion relation $\epsilon_{\vec{k},\sigma}$ can be independently determined. The required momentum resolution can be obtained by time-of-flight or band-mapping techniques. In quantum gas microscopes with a restricted field of view, a $T/4$ oscillation in a harmonic trapping potential \cite{Murthy2014} can be utilized to map momentum to position states \cite{Bohrdt2018,Brown2019}. A detailed discussion of the proposed ARPES scheme can be found in Ref.~\cite{Bohrdt2018}.

If no momentum resolution is desired, the separate detection layer can be replaced by a local probe. In this case the $\a$-particles have no dispersion: hence their final energy is determined by $\epsilon_{\vec{k},\sigma} \equiv \Delta$ and momentum resolution is no longer required to obtain the final state energy. An experimental realization requires a localized optical potential to spatially confine the final $\a$-state \cite{Kollath2007}. 

In the two-layer setting, the ARPES coupling itself, Eq.~\eqref{eqDefVA}, can be realized by introducing a weak tunnel coupling $\delta t_{\rm A}$ between the layers and modulating it at the frequency $\omega_{\rm A}$. This can be easily achieved by an intensity modulation of the lattice beams. In settings where an auxiliary internal atomic state is used, the ARPES coupling corresponds to a radio-frequency, or microwave, transition with frequency $\omega_{\rm A}$.

Finally the lattice modulation, Eq.~\eqref{eqDefVL}, with phase choices as in Eqs.~\eqref{eqPhaseMod1}, \eqref{eqPhaseMod2} can be realized by modulated super-lattice potentials in $x$- and $y$-direction. The cases of $s$- and $d$-wave modulations are particularly simple and do not even require an extra super-lattice potential: Here it is sufficient to homogeneously modulate the tunnelings along $x$- and $y$-directions. For $s$-wave, modulations in phase are needed, while $d$-wave requires a $\pi$-phase shift between the modulations along $x$ and $y$ respectively.

\subsubsection{Realizations in solids}
In copper-oxide layers, we propose to combine state-of-the-art ARPES with a periodically driven optical phonon mode. These phonons couple to the hopping integral of the electrons, thus realizing the desired lattice modulation $\hat{V}_{\rm L}(t)$. The symmetry of the involved phonon mode determines the phases $\phi_{\ij}$ of this lattice modulation, responsible for the transfer of angular momentum to the many-body system. 

In the following we will only consider two distinct Raman-active phonon modes, with $s$- and $d$-wave symmetry, respectively. Specifically we will discuss the $A_{1g}$ and $B_{1g}$ buckling modes in the YBCO class. Because the Cu-O plane is not the symmetry plane in these materials, a crystal electric field can introduce a linear coupling of the electrons to the buckling modes of the oxygen ions \cite{Devereaux1994,Rosch2004}. The effect of such phonons in the effective $t-J$ model of the cuprates is to modify the tunneling matrix elements as \cite{Normand1995}:
\begin{equation}
 t_{\langle \vec{j}, \vec{j} + \vec{e}_\mu \rangle} = t  \left[ 1 - \lambda_t ( u_{\vec{j}}^\mu / a ) \right] , \qquad \mu=x,y;
 \label{eqTModulation}
\end{equation}
see Ref.~\cite{Devereaux1994} for a discussion of the effect in the three-band model. Here $\vec{e}_\mu$ denotes the unit vector along $\mu$-direction, $a$ is the lattice constant, $\lambda_t$ is a dimensionless phonon coupling, and $u_{\vec{j}}^x$ [$u_{\vec{j}}^y$] is the displacement of the $O(2)$ [$O(3)$] oxygen ion from its equilibrium position along the $c$ axis of the crystal. Note that a similar modification as in Eq.~\eqref{eqTModulation} is expected for the exchange integral \cite{Normand1995,Sherman1997}, which we neglect here. 

The symmetry of the phonon mode determines the relative signs of $u_{\vec{j}}^\mu$: For the $A_{1g}$ mode with $s$-wave symmetry, $u_{\vec{j}}^x = u_{\vec{j}}^y$; the $B_{1g}$ mode has $d$-wave symmetry:
\begin{equation}
u_{\vec{j}}^x = - u_{\vec{j}}^y,
\end{equation}
which is the central ingredient required for measuring the $d$-wave rotational ARPES spectrum. 

Combining the ingredients above, we can use the following electron-phonon Hamiltonian to describe the lattice modulation \cite{Rosch2004}:
\begin{equation}
 \hat{V}_{\rm L} =   \sum_{\ij, \sigma}  \cd_{\vec{j},\sigma} \c_{\vec{i},\sigma} \sum_{\vec{q},\nu} g_{\ij}(\vec{q},\nu) \l \b_{\vec{q},\nu} + \bd_{-\vec{q},\nu}  \r. 
 \label{eqHeffVLsolids}
\end{equation}
Here $\nu$ labels the relevant phonon modes and the coupling $g_{\ij}(\vec{q},\nu) \propto t \lambda_t$ reflects the symmetry of the respective phonon mode. 

To measure the rotational ARPES spectrum, we propose to drive the desired $\vec{q}=0$ phonon using a pair of Raman lasers, see inset in Fig.~\ref{figRARPES}; Thus our scheme becomes a three-photon setup. Effectively we can describe the Raman drive by replacing $\b_{\vec{q},\nu} + \bd_{-\vec{q},\nu}$ in Eq.~\eqref{eqHeffVLsolids} with $\delta_{\vec{q},0} \beta_\nu e^{-i \omega_{\rm L} t} /2 + \hc $, where $\omega_{\rm L}$ is the frequency of the drive. To obtain a sizable amplitude $\beta_\nu$, the frequency $\omega_{\rm L}$ should be close to the phonon frequency $\Omega_\nu$. For example, the $B_{1g}$ mode in YBCO is located at $\Omega_{B_{1g}} \approx 42 {\rm meV}$. As a result, we obtain $ \hat{V}_{\rm L}(t) $ as in Eq.~\eqref{eqDefVL} with:
\begin{equation}
 \delta t_{\rm L} = | \beta_\nu ~ g_{\ij}(\vec{0}, \nu)|,~~~ \phi_{\ij} = \begin{cases} 
 e^{i 2 \varphi_{\vec{i} - \vec{j}}}, & \nu = B_{1g} \\ 
1,  & \nu = A_{1g} \end{cases} .
\end{equation}

Since typical super-exchange energies in YBCO materials are around $J \approx 250 {\rm meV}$, we obtain a situation where the ARPES beam is essentially resonant and can directly create hole excitations. I.e. the detuning between the two Raman beams should be around $\omega_{\rm L} \approx 40 {\rm meV}$. In this regime, we use Eq.~\eqref{eqGammaApprx} and assume that the lifetime of hole excitations is short. This is justified by the overall broad structure of the conventional ARPES spectrum, see e.g. dash-dotted bottom line in Fig.~\ref{figRotSetup} (c). To distinguish the direct and rotational ARPES signals, we propose to take a difference measurement of the photo-electron signal with ($\delta t_{\rm L} \neq 0$) and without ($\delta t_{\rm L} = 0$) the Raman beams on.

\subsubsection{Extensions to STM probes}
The general multi-photon scheme proposed above to probe rotational excitations of mobile dopants can be straightforwardly generalized to scanning-tunneling microscopy (STM) setups. In this case the photo-electron with momentum $\vec{k}$ created by $\hat{V}_{\rm A}$ is replaced by a localized electronic final state at site $\vec{j}_0$, i.e. 
\begin{equation}
 \hat{V}_{\rm A, STM}(t) = - \delta t_{\rm A} \sin( \omega_{\rm A} t)  \sum_{\sigma} \l  \ad_{\vec{j}_0,\sigma} \c_{\vec{j}_0,\sigma} + \hc \r.
 \label{eqDefVAstm}
\end{equation}
The resulting rotational STM signal is obtained by integrating the rotational ARPES over all momenta:
\begin{equation}
I_{\rm rot}^{(m)}(\omega) = \int_{\rm BZ} d^2\vec{k} ~ A_{\rm rot}^{(m)}(\vec{k}, \omega).
\end{equation}

\begin{figure}[t!]
\centering
\epsfig{file=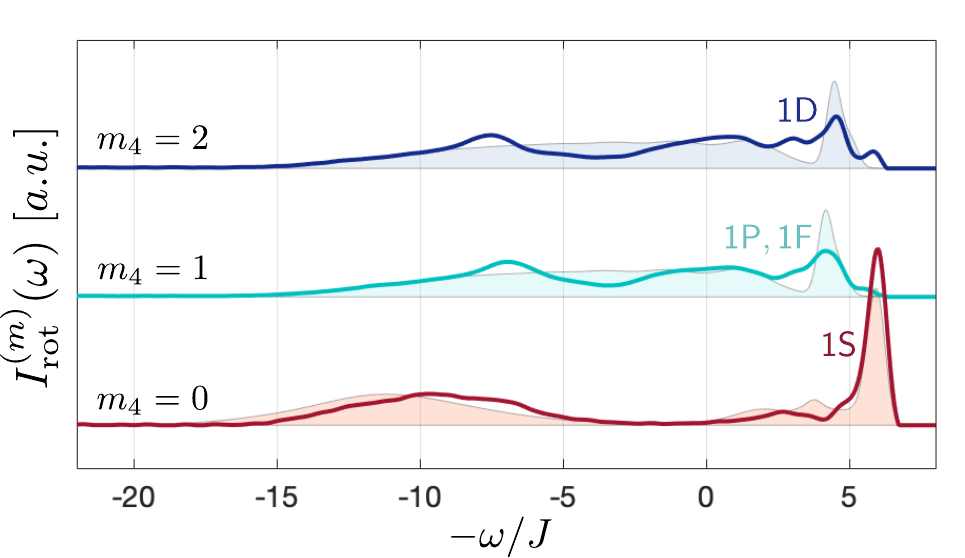, width=0.49 \textwidth}
\caption{\textbf{Rotational STM spectra.} The integrated rotational ARPES signal $I^{(m)}_{\rm rot}(\omega)$ which can be measured by the proposed multi-photon rotational STM scheme is calculated for the $t-J$ model at $t/J=3$. The rotational meson states are still clearly visible as pronounced peaks above the ground state. We consider an extended $L_x=40\times 4$-leg cylinder, for which the rotational ARPES signal is integrated over $k_x$ and summed over all discrete values of $k_y=-\pi/2,0,\pi/2,\pi$. We compare our numerical DMRG results (solid lines) with predictions by the spinon-chargon toy model (shaded areas); for the latter, the four lowest lying states were broadened by $J/4$ (similar to the observed Fourier broadening in the DMRG numerics) while all higher excited states were artificially broadened by $J$. In addition the toy model curves for $m_4=1$ ($m_4=2$) were shifted by a fitted $\Delta \omega = 0.9 J$ ($\Delta \omega = 1.2 J$) towards the lowest-lying quasiparticle peak. 
}
\label{figSTMrot}
\end{figure}

The bandwidth of rotational excitations (with $m_4 \neq 0$) is significantly smaller than in the vibrational ground state (with $m_4=0$). Hence the rotational STM protocol without full momentum resolution is sufficient to resolve rotational meson excitations. As an example, we show the expected STM signal for the experimentally most relevant case $t/J=3$ in Fig.~\ref{figSTMrot}. Indeed, in addition to the most prominent vibrational ground state at $m_4=0$, the rotational excitations at $m_4=1,2$ are clearly visible as distinct quasiparticle peaks. For $m_4=1,2$ we also observe a weak signal at the ground state energy, owing to hybridization of rotational and vibrational states at non-C4IM. 

In Fig.~\ref{figSTMrot} we also compare our numerical results to the expected integrated spectrum from the spinon-chargon toy model. In the latter we include small shifts on the energy axis and weak broadening as fit parameters to obtain better agreement. As for the ARPES spectra at the nodal point, shown in the main text, we find that the toy model reasonably predicts the overall shape even of the incoherent part of the spectrum and the strong suppression of spectral weight at high energies.

\subsection{TD-DMRG simulations}
\label{MethodsB}
We use time-dependent matrix product state methods \cite{Schollwock2011,Kjall2013,Zaletel2015,Paeckel2019a}, in particular the TeNPy package \cite{Hauschild2018SciPost,hauschildTenpy}, to numerically calculate the rotational ARPES spectrum introduced in the main text. To this end we start from the numerically obtained ground state $\ket{\Psi_0}$ of the Heisenberg model on a $4 \times L_x$ cylinder, and apply the rotational operator $\hat{R}_{m,\sigma}(\vec{j})$ introduced in Eq.~\eqref{eqDefRmj} below.
We assume that the resulting one-hole states are described by the $t-J$ model on the same lattice. In two dimensions and for a single hole, the Hamiltonian of the $t-J$ model becomes~\cite{Auerbach1998},
\begin{equation}
\H_{t-J} =  -t \sum_{\ij, \sigma} \hat{\mathcal{P}} \left( \cd_{\vec{i},\sigma} \c_{\vec{j},\sigma} + h.c. \right) 
\hat{\mathcal{P}}+ J \sum_\ij \hat{\mathbf{S}}_{\vec{i}} \cdot \hat{\mathbf{S}}_{\vec{j}},
\label{eq:tjmodel}
\end{equation}
where $\hat{\mathcal{P}}$ projects on states with less than two fermions $\c^{(\dagger)}_{\vec{j},\sigma}$ per site. The first term describes tunneling of holes with amplitude $t$ and the second term denotes spin-exchange interactions with coupling constant $J=4 t^2/U$, where $U$ is the Hubbard interaction. 

To evaluate the rotational spectral function $ A_{\rm rot}^{(m)}(\vec{k},\omega)$ from Eq.~\eqref{eqDefArot}, we express it in real space and time,
\begin{equation}
 A_{\rm rot}^{(m)}(\vec{k},\omega)  = \int_0^{\infty} dt \sum_{\vec{j}} e^{-i \vec{k} \cdot \vec{j}} \mathcal{G}_{\rm rot}^{(m_4)}(\vec{j},t),
 \label{eqArotTime}
\end{equation}
see also Ref.~\cite{Bohrdt2020}, where the rotational Green's function in real space is defined as 
\begin{equation}
\mathcal{G}_{\rm rot}^{(m_4)}(\vec{j},t) = \theta(t) \sum_\sigma \bra{\Psi_0} \hat{R}^\dagger_{m_4,\sigma}(\vec{j},t) \hat{R}_{m_4,\sigma}(\vec{0},0) \ket{\Psi_0},
\end{equation}
with the corresponding rotational operator
\begin{equation}
\hat{R}_{m_4,\sigma}(\vec{j}) =   \sum_{\vec{i}: \ij} e^{i m_4 \varphi_{\vec{i} - \vec{j}}}  \sum_{\sigma'} \cd_{\vec{j},\sigma'} \c_{\vec{i},\sigma'} \c_{\vec{j},\sigma}.
\label{eqDefRmj}
\end{equation}

We start by numerically calculating the ground state of the Heisenberg model, using a bond dimension of $\chi=600$. Subsequently, we apply the rotational operator as defined in Eq.~\eqref{eqDefRmj} in the origin and time-evolve the resulting state under the $t-J$ Hamiltonian using time-dependent matrix product state methods \cite{Kjall2013,Zaletel2015}. We thus obtain the rotational Green's function in real space and time. We use linear prediction to increase the time window and multiply our data with a Gaussian envelope in order to minimize the weight of the data generated by said linear prediction in the spectrum \cite{Verresen2018,Bohrdt2020}. Finally, we perform a Fourier transformation in time and space to obtain the spectral function $ A_{\rm rot}^{(m)}(\vec{k},\omega)$ from Eq.~\eqref{eqDefArot}. We carefully checked our results for convergence with the bond dimension.

\subsection{Spinon-chargon toy model}
\label{MethodsC}
The identification of rotational and vibrational resonances in the spectrum provide compelling evidence that magnetic polarons are composite objects with an internal structure. We describe the latter by an effective theory, which models magnetic polarons as bound states of spinons and chargons connected by a string on a square lattice, see Fig.~\ref{figRotSetup} (a), with a linear string tension calculated from spin-correlations in the undoped parent AFM \cite{Grusdt2018PRX}. In addition, we extend earlier approaches \cite{Bulaevskii1968,Brinkman1970,Trugman1988,Shraiman1988a,Grusdt2018PRX,Grusdt2019PRB} by including spinon dynamics explicitly in our theory. Details of our theoretical description are presented in the supplements \ref{SuppMatToyModel}.

In Fig.~\ref{figRotSetup} (c) we compare our DMRG spectra to predictions by the spinon-chargon toy model. To capture Fourier broadening present in our DMRG results, we broadened the lowest rotational and vibrational peaks in the toy model by $\sigma_0=J/4$. For a better comparison of the overall spectral weight and shape, we added small overall energy shifts $\Delta \omega$ separately for each $m_4$ and introduced broadening $\sigma_1 = J$ of all higher excited states. Such broadening is expected to arise from couplings to magnon excitations in the AFM \cite{Grusdt2018PRX,Wrzosek2020} which we neglect in our toy model calculation so far. 

The resulting toy-model prediction is in good agreement with the full numerical spectra: It captures the spectral weight of the low-energy mesonic resonances $\mathsf{1S}$, $\mathsf{2S}$, $\mathsf{1P}$, $\mathsf{1D}$ and $\mathsf{1F}$. Remarkably, even the spectral features at higher energies are correctly described. In particular this includes the strong suppression of spectral weight in the $m_4=0$ rotational spectrum (red line in Fig.~\ref{figRotSetup}) between $- \omega = -5 J$ to $\omega = 0$, which is followed by a broad continuum at higher energies. This should be contrasted with the non-vanishing spectral weight found in the same frequency range for $m_4 \neq 0$ rotational spectra and for the standard ARPES spectrum with $m_4=0$ (bottom line in Fig.~\ref{figRotSetup}). The tails at the highest energies are also correctly described by the toy model.

In Fig.~\ref{figReggeTraj} we compare Regge-like trajectories. Here we used the simplified toy model \cite{Grusdt2018PRX} (solid lines in Fig.~\ref{figReggeTraj}) which neglects spinon dynamics and explains the characteristic scaling with $J/t$, without any free fit parameters. A comparison to the full spinon-chargon toy model including spinon dynamics is provided in Fig.~\ref{figToyModelRegge} in the supplements (section \ref{SecToyModelRes}): There we confirm the power-laws from Eq.~\eqref{eqErotScaling}, \eqref{eqEvibScaling} and find similar quantitative agreement as in Fig.~\ref{figReggeTraj}, again without any free fit parameters. Notably, the refined spinon-chargon toy model with spinon dynamics predicts a weak splitting of $\mathsf{1P}$/$\mathsf{1F}$ and $\mathsf{1D}$ resonances as found by DMRG. Further comparison of Regge-like trajectories to the full spinon-chargon toy model can be found in the supplements \ref{SuppMatAddNum}.

In the insets in Fig.~\ref{figSpectralLines} (b) and (c) we show full spectral cuts at low energy predicted by the spinon-chargon toy model, without any free fit parameters. As observed in our full DMRG results, we find that the rotational mesonic states have much weaker dispersion than the $\mathsf{1S}$ magnetic polaron ground state. The toy model predicts some entirely flat bands, similar to the flat bands predicted for mesonic bound states of two identical partons \cite{Shraiman1988a}, in addition to weakly dispersing bands. The overall distribution of spectral weight is also qualitatively captured by the toy model.

\section*{Acknowledgements}
The authors thank M. Knap, Z.X. Shen, A. Cavalleri, E.-A. Kim, I. Morera Navarro, S. Sachdev, U. Schollw\"ock, I. Bloch, M. Greiner, J. Koepsell, and F. Pollmann for fruitful discussions.
This research was supported by the Deutsche Forschungsgemeinschaft (DFG, German Research Foundation) under Germany's Excellence Strategy -- EXC-2111 -- 390814868. ED was supported by the ARO grant number W911NF-20-1-0163, the NSF grant EAGER-QAC-QSA 2222-206-2014111, the NSF grant OAC-1934714, and the Harvard-MIT CUA.

\section*{References}

%

\onecolumngrid
\newpage

\section{Supplementary Material}
\setcounter{subsection}{0}

\subsection{Theory of spinon-chargon bound states}
\label{SuppMatToyModel}
Here we present the string-based model of spinon-chargon bound states in a 2D AFM. To include the momentum dependence of the bound states, we improve previous theoretical models based on geometric strings \cite{Grusdt2018PRX,Grusdt2019PRB} by including spinon dynamics beyond the strong-coupling limit. To this end we work in the co-moving frame with the spinon.

\subsubsection{Model}
We include strong spin-charge correlations by working in the effective Hilbert space obtained by the geometric string construction \cite{Grusdt2018PRX}. The corresponding basis states are labeled by the position of the spinon $\vec{x}_{\rm s}$ in the 2D lattice, and the string $\Sigma$ along which spins are displaced. The chargon (spinon) is located at the end (beginning) of the string $\Sigma$. Here $\Sigma=\{ \vec{e}_1, \vec{e}_2,...,\vec{e}_{\ell} \}$ denotes a sequence of steps $\vec{e}_n = \pm \vec{e}_{x,y}$ without direct re-tracing, i.e. $\vec{e}_{n+1} \neq - \vec{e}_n$; more conveniently, string states $\Sigma$ can be represented by the sites of a Bethe-lattice, or Cayley tree, with coordination number $z=4$. 

Every spinon-chargon basis state $\ket{\vec{x}_{\rm s}, \Sigma}$ has a microscopic representation by a quantum state $\ket{\psi(\vec{x}_{\rm s}, \Sigma)}$ in the $t-J$ model, defined by
\begin{equation}
\ket{\psi(\vec{x}_{\rm s}, \Sigma)} = \hat{G}_\Sigma \sum_\sigma \c_{\vec{x}_{\rm s}, \sigma} \ket{\Psi_0},
\end{equation}
where $\c_{\vec{j},\sigma}$ is a microscopic fermion operator at site $\vec{j}$ with spin $\sigma$. Further, $\ket{\Psi_0}$ denotes the ground state of the undoped Heisenberg model and the operator $\hat{G}_\Sigma$ displaces all spins along the string $\Sigma$ while simultaneously moving the hole \cite{Grusdt2019PRB}. 

The geometric string states $\ket{\psi(\vec{x}_{\rm s}, \Sigma)}$ form an over-complete and non-orthogonal basis of the one-hole $t-J$ Hilbert space. However, to a good approximation we may assume that most of the relevant string states are orthonormal \cite{Grusdt2019PRB}. This motivates our definition of the effective spinon-chargon Hilbertspace, which is spanned by the set of orthonormal basis states $\ket{\vec{x}_{\rm s}, \Sigma}$ with
\begin{equation}
 \langle  \vec{x}_{\rm s}, \Sigma \ket{ \vec{x}_{\rm s}' , \Sigma'} = \delta_{\Sigma,\Sigma'} \delta_{\vec{x}_{\rm s} , \vec{x}_{\rm s}' }.
\end{equation}
Note that our choice of the Hilbert space is similar to the non-retracing string approximation proposed by Brinkman and Rice \cite{Brinkman1970}, but in addition we include the spinon degrees of freedom $\vec{x}_{\rm s}$. 

The effective Hamiltonian $\H$ describing spinon-chargon bound states can be obtained by calculating matrix elements of the microscopic $t-J$ Hamiltonian, $\bra{\psi(\vec{x}_{\rm s}, \Sigma)} \H_{tJ} \ket{\psi(\vec{x}_{\rm s}', \Sigma')}$. The hopping part $\propto t$ yields tunnelings between nearest-neighbor sites $\langle \Sigma, \Sigma' \rangle$ on the Bethe lattice:
\begin{equation}
\H_t^{\rm c} = - t \sum_{\langle \Sigma, \Sigma' \rangle} \ket{\Sigma} \bra{\Sigma'} + \hc,
\end{equation}
independent of the spinon position. 

The spin-exchange terms $\propto J_\perp$ in $\H_{tJ}$ introduce spinon dynamics. Assuming for simplicity that $\ket{\Psi_0}$ is given by a classical N\'eel state along $z$, we obtain next-nearest neighbor tunneling of the spinon \cite{Grusdt2019PRB} which is correlated with a re-organization of the string:
\begin{equation}
 \H_{J}^{\rm s} = \frac{J_\perp}{2} \sum_{\vec{x}_{\rm s},\Sigma} \sum_{(\vec{e}_2,\vec{e}_1)} \!\! {\vphantom{\sum}}' \ket{\vec{x}_{\rm s} + \vec{e}_2 + \vec{e}_1} \bra{\vec{x}_{\rm s}}   \otimes  \ket{\Sigma_{\vec{e}_2,\vec{e}_1}} \bra{\Sigma}.
 \label{eqHJs}
\end{equation}
Here the sum $\Sigma'$ is over consecutive links $\vec{e}_2 \neq - \vec{e}_1$ for which the spins on sites $\vec{x}_{\rm s}+\vec{e}_1$ and $\vec{x}_{\rm s}+\vec{e}_1 + \vec{e}_2$ are anti-aligned for the given string configuration $\Sigma$; the string $\Sigma_{\vec{e}_2,\vec{e}_1}$ is obtained by adding or removing the first two steps $\vec{e}_1$ and $\vec{e}_2$ from the original string $\Sigma$ (see Ref.~\cite{Grusdt2019PRB} for a discussion).

The remaining spin-exchange terms $\propto J_{z}, J_\perp$ in $\H_{tJ}$ give rise to spinon-chargon interactions,
\begin{equation}
 \H_J^{\rm sc} = \sum_\Sigma V_\Sigma \ket{\Sigma} \bra{\Sigma} \approx  \sum_\Sigma V(\ell_\Sigma) \ket{\Sigma} \bra{\Sigma}.
 \label{eqHJsc}
\end{equation} 
In the second step we assume that the string potential $V_\Sigma$ depends only on the length of the string $\ell_\Sigma$ (linear string approximation). We calculate $V(\ell_\Sigma)$ by considering straight strings, which yields 
\begin{equation}
V(\ell_\Sigma) = \frac{dE}{d\ell} \ell_\Sigma + g_0 \delta_{\ell_\Sigma,0} + \mu_{\rm h}.
\label{eqVsgmaLinear}
\end{equation}
The linear string tension $dE/d\ell$, $g_0$ and $\mu_{\rm h}$ can be expressed in terms of the correlations in the undoped parent AFM, see \cite{Grusdt2019PRB}.

\subsubsection{Symmetries and quantum numbers}
The effective Hamiltonian defined in the spinon-chargon Hilbert space,
\begin{equation}
 \H = \H_t^{\rm c}  +  \H_{J}^{\rm s} +  \H_J^{\rm sc},
 \label{eqDefHeff}
\end{equation}
is manifestly translationally invariant, $[\H , \hat{T}_\mu]=0$. The translation operator leaves the string state unchanged and shifts the spinon position, $\hat{T}_\mu = e^{- i \hat{\vec{P}}_{\rm s} \cdot \vec{e}_\mu}$ where $\mu = x,y$. Hence, we can label eigenstates by the spinon momentum $\vec{k}_{\rm s}$ in the lattice. 

Furthermore, the underlying $t-J$ model has an exact $\hat{C}_4$ discrete rotational symmetry, which carries over to the effective Hamiltonian: $[\H, \hat{C}_4]=0$. In the new Hilbert space, $\hat{C}_4$ rotates the spinon position $\vec{x}_{\rm s}$ (the string configuration $\Sigma$) around the origin in the lattice (of the Bethe lattice). Hence, eigenstates can also be labeled by  $m_4=0,1,2,3$ corresponding to eigenvalues $\exp( i m_4 \nicefrac{\pi}{2})$ of $\hat{C}_4$. 

In general, $\hat{C}_4$ and $\hat{T}_\mu$ do note commute and we cannot simultaneously assign linear and angular momentum quantum numbers. Exceptions require momenta $\vec{k}_{\rm s}$ for which $\H(\vec{C}_4 \vec{k}_{\rm s}) = \H(\vec{k}_{\rm s})$. In particular, this is the case for $C_4$-invariant momenta (C4IM), i.e. when the rotated momentum $\vec{C}_4 \vec{k}_{\rm s}$ is equivalent to $\vec{k}_{\rm s}$ modulo the reciprocal lattice vector, 
\begin{equation}
\vec{C}_4 \vec{k}_{\rm s}^{\rm C4IM} \equiv \vec{k}_{\rm s}^{\rm C4IM} {\rm mod} \vec{G}.
\end{equation}

The reciprocal lattice vectors depend on the unit cell. In the square lattice with a one-site unit-cell, the resulting C4IM are: $\vec{k}_{\rm s}^{\rm C4IM} = (0,0)$ and $(\pi,\pi)$. In the AFM phase, where the sub-lattice symmetry is spontaneously broken, the C4IM of the magnetic unit-cell are
\begin{equation}
\vec{k}_{\rm s}^{\rm C4IM} = (0,0), (\pi,0), (0,\pi), (\pi,\pi).
\label{eqC4IMmbz}
\end{equation}
While the anti-nodal points are $C_4$ invariant, we emphasize that the nodal points $(\pm \pi/2, \pm \pi/2)$, where the ground state of the magnetic polarons is located, are \emph{not} $C_4$ invariant.

If we make the linear string approximation in Eq.~\eqref{eqHJsc}, the system at strong coupling $t \gg J_\perp$ has a series of additional discrete $\hat{C}_3$ rotational symmetries around the nodes of the Bethe lattice different from the origin; see Ref.~\cite{Grusdt2018PRX} for a detailed discussion of the resulting $m_3$ eigenvalues.  

\subsubsection{Solution within linear string approximation}
To solve the effective Hamiltonian \eqref{eqDefHeff} for the linear string potential \eqref{eqVsgmaLinear}, we start from the following basis,
\begin{equation}
 \ket{\vec{k}_{\rm s}, \ell_\Sigma, \vec{m}} = \frac{1}{\sqrt{V}} \sum_{\vec{x}_{\rm s}} e^{i \vec{k}_{\rm s} \cdot \vec{x}_{\rm s}} \ket{\vec{x}_{\rm s}, \ell_\Sigma, \vec{m}}.
\end{equation}
Here $\ell_\Sigma$ and $\vec{m}=(m_4,m_3^{(1)},m_3^{(2)},...)$ denote the string length and angular momenta on the Bethe lattice to label string configurations $\Sigma$ \cite{Grusdt2018PRX}; $V = L^2$ is the area of the physical lattice.

\begin{figure}[t!]
\centering
\epsfig{file=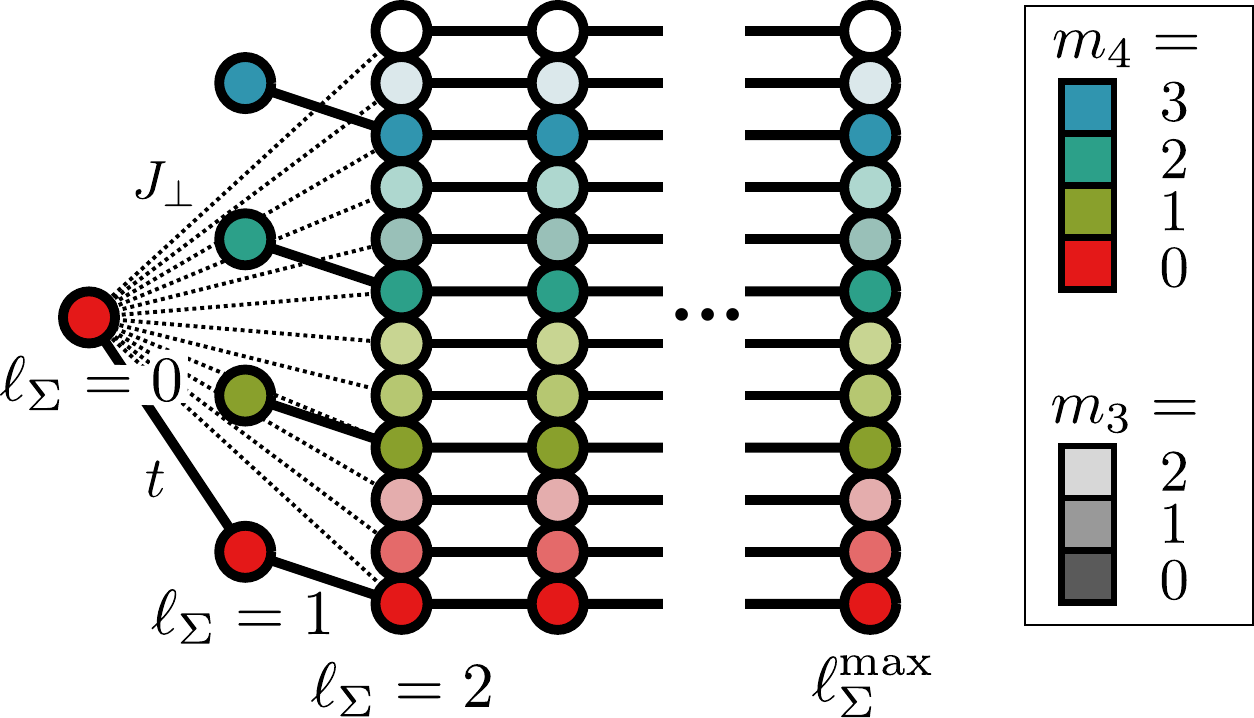, width=0.4 \textwidth}
\caption{\textbf{Truncated rotational string basis.} For fixed center-of-mass momentum $\vec{k}_{\rm s}$ of the spinon-chargon pair, we work with a truncated string basis. At string lengths $\ell_\Sigma=0,1$ all states are included, and labeled by their discrete $C_4$ and $C_3$ angular momentum quantum numbers $m_4$ and $m_3=m_3^{(1)}$ in the Bethe lattice. For longer strings with length $\geq 3$ we only include $m_4$ and $m_3$ excitations and set higher $m_3^{(>1)} = 0$. The generally allowed matrix elements for hole (solid lines, $t$) and spinon (dotted lines, $J_\perp$) hopping are indicated (in the case of the spinon, only those involving $\ell_\Sigma=0$, for clarity). 
}
\label{figTruncatedBasis}
\end{figure}

We truncate the basis by neglecting higher angular momenta beyond $m_3 \equiv m_3^{(1)}$; i.e. we consider only states with $m_3^{(n)}=0$ for $n \geq 2$, see Fig.~\ref{figTruncatedBasis}. This is motivated by the strong coupling result ($J_\perp=0$) that non-zero rotational quantum numbers $m_3^{(n)} \neq 0$ lead to higher spinon-chargon interaction energies \cite{Grusdt2018PRX}: This is easily understood by noting that non-zero values $m_3^{(n)} \neq 0$ correspond to eigenvectors which are superpositions of strings with lengths $\geq n+1$, see Ref.~\cite{Grusdt2018PRX}.

Since $\vec{k}_{\rm s}$ is conserved, the effective Hamiltonian in the truncated basis is fully defined by its matrix elements
\begin{equation}
 H_{\ell', m_4', m_3' ; \ell, m_4, m_3}(\vec{k}_{\rm s}) = \bra{\vec{k}_{\rm s}, \ell', \underbrace{m_4', m_3'}_{\vec{m}'}} \H \ket{ \vec{k}_{\rm s}, \ell, \underbrace{m_4, m_3}_{\vec{m}} }.
\end{equation}
The latter are relatively easy to calculate if we make use of symmetries: The chargon hopping $\H_t^{\rm c}$ conserves all angular momenta $\vec{m}$ on the Bethe lattice \cite{Grusdt2018PRX} and couples only $\ell$ and $\ell \pm 1$. It is sufficient to consider one direction -- we choose $\ell \to \ell' = \ell-1$ -- since the other follows from the condition that $\H$ is hermitian. From $\H_t^{\rm c}$ we obtain:
\begin{equation}
H^{\rm c}_{\ell-1, \vec{m}' ; \ell, \vec{m}} = - t \delta_{\vec{m}',\vec{m}}  \times \begin{cases} 
\sqrt{z-1}, &\ell \geq 2  \\ 
\sqrt{z}, &\ell = 1 \end{cases}
\end{equation}
independent of $\vec{k}_{\rm s}$, where $z=4$ is the coordination number of the lattice. 

For the spinon hopping Eq.~\eqref{eqHJs} only transitions between $\ell \to \ell \pm 2$ are allowed and the angular momenta $\vec{m}$ can change in this process. A full calculation for our lattice with $z=4$ yields
\begin{equation}
H^{\rm s}_{\ell-2, \vec{m}' ; \ell, \vec{m}} = J_\perp  \times \begin{cases} 
\frac{1}{\sqrt{3}} [ \Lambda^{\rm s}_{\vec{m}} \delta_{\vec{m}',\vec{0}} - \Phi^{\rm s}_{\vec{m}',\vec{m}} ], &\ell > 2  \\ 
\frac{1}{2} \delta_{\vec{m}',\vec{0}} \Lambda^{\rm s}_{\vec{m}}, &\ell = 2 \end{cases}.
\end{equation}
Here we first defined
\begin{equation}
\Lambda^{\rm s}_{\vec{m}}(\vec{k}_{\rm s}) = \frac{1}{2 \sqrt{3}} \sum_{\nu=0}^3 \sum_{\nu'=1}^3 e^{i \nu m_4 \frac{\pi}{2}} e^{i \nu' m_3 \frac{2 \pi}{3}} e^{- i \vec{k}_{\rm s} \cdot \vec{e}_{\nu',\nu}},
\end{equation}
with $\vec{e}_{\nu',\nu}$ re-tracing the first two string segments, starting to count at the spinon position; $\nu \pi/2$ denotes the angle of the first string segment relative to the $x$-axis (i.e. $\nu=0,1,2,3$) and $(\nu'-2) \pi/2$ denotes the angle of the second string segment relative to the first (i.e. $\nu'=1,2,3$). In complex notation (i.e. real and imaginary parts of $\epsilon_{\nu',\nu} \in \mathbb{C}$ represent the $x$ and $y$ components of $\vec{e}_{\nu',\nu} \in \mathbb{R}$) it holds: 
\begin{equation}
\epsilon_{\nu',\nu} = e^{i \nu \frac{\pi}{2}} + e^{i \nu \frac{\pi}{2}} e^{i (\nu'-2) \frac{\pi}{2}}.
\end{equation}
We further defined for longer strings:
\begin{equation}
 \Phi^{\rm s}_{\vec{m}',\vec{m}}(\vec{k}_{\rm s}) = \frac{1}{4} \sum_{\nu=0}^3 e^{i (m_4 - m_4') \nu \frac{\pi}{2}} \chi^{\rm s}_{\vec{m}}(\vec{k}_{\rm s},\nu),
 \label{eqDefPhiS}
\end{equation}
with:
\begin{equation}
\chi^{\rm s}_{\vec{m}}(\vec{k}_{\rm s},\nu) = \frac{1}{2 \sqrt{3}} \sum_{\nu'=1}^3 e^{i \nu' m_3 \frac{2 \pi}{3}} e^{- i m_4 \nu' \frac{\pi}{2} } e^{- i \vec{k}_{\rm s} \cdot \vec{e}_{\nu',\nu-\nu'}}.
\label{eqDefChiS}
\end{equation}

By diagonalizing the effective Hamiltonian $H^{\rm s}(\vec{k}_{\rm s})$ in the truncated basis (Fig.~\ref{figTruncatedBasis}), we obtain all low-energy spinon-chargon bound states and their dispersion relations. The ground state is adiabatically connected to $\vec{m}=0$ without rotational excitations at $\vec{k}_{\rm s}^{\rm C4IM}$; within our simplified spinon model Eq.~\eqref{eqHJs}, it has a degenerate energy minimum at the edge of the magnetic Brillouin zone including nodal and anti-nodal points. This dispersion closely resembles the ground state magnetic polaron dispersion, although it misses the small energy splitting between nodal and anti-nodal points \cite{Grusdt2019PRB}. The low-energy excited states have non-trivial rotational quantum numbers, and their dispersion relations feature a richer structure. The spinon hopping causes quantum interference effects between rotationally excited states which are degenerate in the absence of spinon hopping.

At the $\vec{k}_{\rm s}^{\rm C4IM}$ of the magnetic Brillouin zone, see Eq.~\eqref{eqC4IMmbz}, the Hamiltonian $H_{\ell', \vec{m}' ; \ell, \vec{m}}(\vec{k}_{\rm s}^{\rm C4IM})$ is block-diagonal, with blocks labeled by $m_4=0,1,2,3$.  At $\vec{k}_{\rm s}=0$ the block with $m_4=0$ also conserves the $m_3$ quantum numbers, since $\chi^{\rm s}_{m_4=0,m_3}(0,\nu) \propto \delta_{m_3,0}$. One further finds that $\chi^{\rm s}_{m_4 \neq 0,m_3=0}(0,\nu)$ is equal for all $m_4=1,2,3$, which means that the three lowest order rotational states with $m_4 \neq 0$, $m_3=0$ are degenerate at $\vec{k}_{\rm s}=0$.

\begin{figure}[b!]
\centering
\epsfig{file=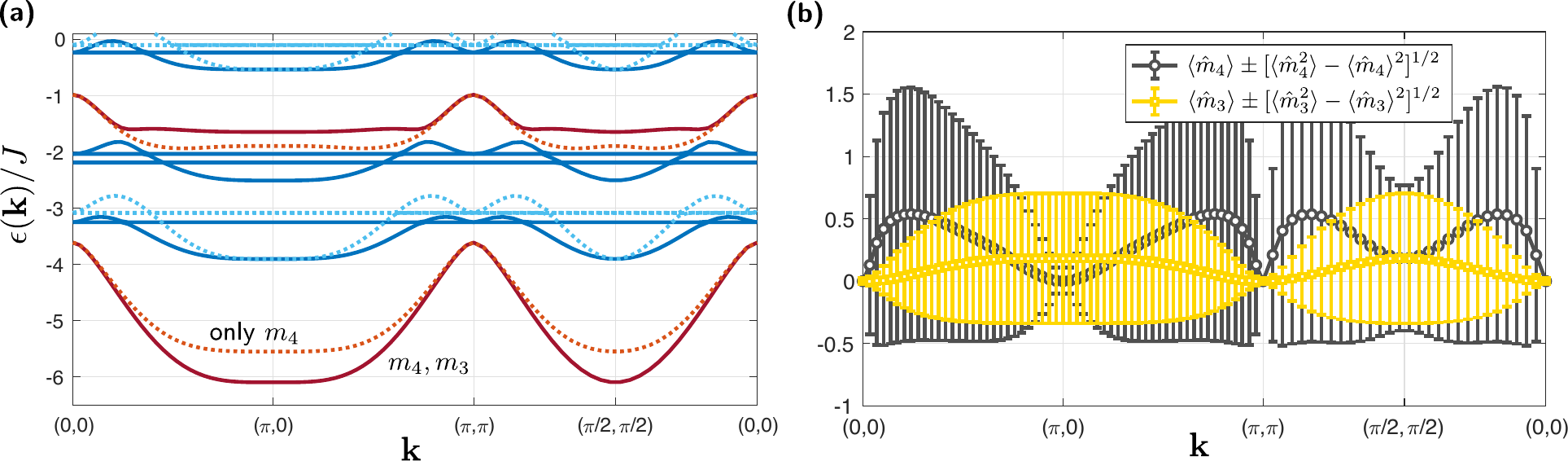, width=0.95 \textwidth}
\caption{\textbf{String toy model results.} (a) We show the lowest energy bands predicted by the spinon-chargon toy model for $t/J=3$ and $J_\perp = J$. A cut along high-symmetry directions in the Brillouin zone is shown. The full solid lines correspond to the full truncated basis; the light dotted lines correspond to a further truncated basis with only $m_4 \neq 0$ included. Colors indicate how states connect to rotational (blue) and vibrational (red) bands at $\vec{k}=\vec{0}$. (b) We show the expectation value and variance (error bars) of the rotational eigenvalues $\hat{m}_4$ (black) and $\hat{m}_3$ (yellow) for the ground state of the spinon-chargon toy model; parameters as in (a). 
}
\label{figToyModelResults}
\end{figure}

\subsubsection{Results}
\label{SecToyModelRes}
Now we apply the spinon-chargon toy model introduced above to calculate the ro-vibrational eigenstates and the spectral function. In Fig.~\ref{figToyModelResults} (a) we show all low-energy spinon-chargon eigenstates along high-symmetry cuts through the Brillouin zone. Although well-defined $C_4$ rotational quantum numbers $m_4$ can only be assigned at C4IM, we can still clearly identify sets of states which are adiabatically connected to the rotational or vibrational states at the C4IM. The ro-vibrational ground state is non-degenerate, lowest red band in Fig.~\ref{figToyModelResults} (a). Then we find a band consisting of three rotational states, which correspond to the non-trivial rotational states $m_4 \neq 0$ at the C4IM, lowest blue band in Fig.~\ref{figToyModelResults} (a). At $\vec{k}=\vec{0} \equiv \vec{\pi} ~ {\rm mod} \vec{G}$, the latter are exactly degenerate. 

To estimate the effect of higher rotational excitations with $m_3^{(n)} \neq 0$, in Fig.~\ref{figToyModelResults} (a) we also compare our results to toy model calculations where we truncate the basis further and include only $m_4$ states while setting all $m_3^{(n)}=0$. For the lowest lying vibrational and rotational states, we observe a modest shift to lower energies when $m_3^{(1)}\neq 0$ states are included. The ground state at $\vec{k}=\vec{0} \equiv \vec{\pi} ~ {\rm mod} \vec{G}$ is an exception: As described in the previous section, all $m_3^{(n)} = 0$ quantum numbers are explicitly conserved at this point in the toy model, and the ground state energy is exactly obtained.

We find that the third band of states (solid blue) we identify in Fig.~\ref{figToyModelResults} (a) consists of eight states, some of which are degenerate. This band is only obtained if $m_3^{(1)}$ excitations are included. Indeed, this number of states was predicted at strong coupling for higher-order rotational excitations with $m_3^{(1)}=1,2$ (each of those has four distinct $m_4$ states)  \cite{Grusdt2018PRX}. Away from the C4IM, the non-trivial $m_3^{(1)}\neq 0$ excitations weakly hybridize with the purely vibrationally excited state ($\mathsf{2S}$) and we observe small avoided crossings. The counting suggests that the energetically highest shown three states correspond to the rotationally excited, $m_4 \neq 0$, versions of the $\mathsf{2S}$ state, with a vibrational quantum number $n=2$.

In Fig.~\ref{figToyModelResults} (b) we calculate the expectation values $\langle \hat{m}_4 \rangle$ and $\langle \hat{m}_3 \rangle$ for the ground state. The error bars denote the variance. As expected, we find that $m_4=0$ is a good quantum number (zero variance) at C4IM of the magnetic Brillouin zone. At $\vec{k}=\vec{0} \equiv \vec{\pi} ~ {\rm mod} \vec{G}$, even $m_3=0$ is a good quantum number with zero fluctuations. All other momenta show some hybridization of $m_4$ and $m_3$ quantum numbers. 

\begin{figure}[t!]
\centering
\epsfig{file=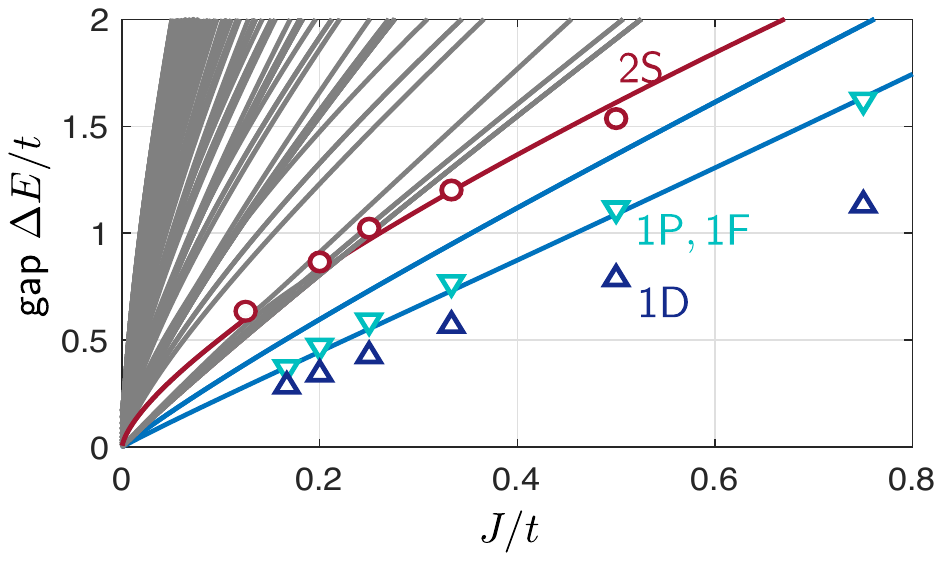, width=0.44 \textwidth}
\caption{\textbf{Meson Regge trajectories from the toy model.} We calculate the excitation energy gaps $\Delta E$ from the ground state from the spinon-chargon toy model at the nodal point, $\vec{k}=(\pi/2,\pi/2)$, solid lines. These predictions are compared to our numerical DMRG results (data points). The lowest excitations can be identified as rotational (blue) and vibrational (red) by the dependence of their energy gap on $J/t$. Higher excited states (gray) show similar, though less pronounced behavior. 
}
\label{figToyModelRegge}
\end{figure}

In Fig.~\ref{figToyModelRegge} we apply the string toy model to calculate Regge trajectories at the nodal point. We find that the energy gap $\Delta E$ to the three lowest-lying excitations scales linearly with $J$, the hallmark signature expected of rotational states. We also compare our results to the numerical DMRG data shown already in the main text, see Fig.~\ref{figReggeTraj}. Without any free fit parameters, we find that the energy gap to the first vibrational excitation ($\mathsf{2S}$) is accurately predicted by the spinon-chargon toy model. 

Because of the hybridization of different $m_3$ and $m_4$ states with each other, the toy model predicts a splitting between different states from the lowest rotational excitation. While the overall scale of this splitting is correctly predicted, we find numerically from DMRG a smaller than expected energy gap to the rotational states. As the DMRG data, the toy model predicts a non-degenerate lower rotational line and a two-fold degenerate higher rotational line. However, we found that the distribution of spectral weight in the toy model differs from the DMRG results. 

In summary, we find that the spinon-chargon toy model correctly captures most qualitative properties of the ro-vibrational meson states that we observe numerically in the $t-J$ model. We expect that additional couplings to magnons, which are neglected in our toy model, can give rise to some of the observed discrepancies. Whether all lines can be fully quantitatively captured by such an approach remains to be seen.

\subsection{Additional numerical results}
\label{SuppMatAddNum}
Here we present additional numerical data supporting our main finding, namely that pronounced rotational quasiparticle peaks exist in the lightly doped $t-J$ model.

\begin{figure*}[t!]
\centering
\epsfig{file=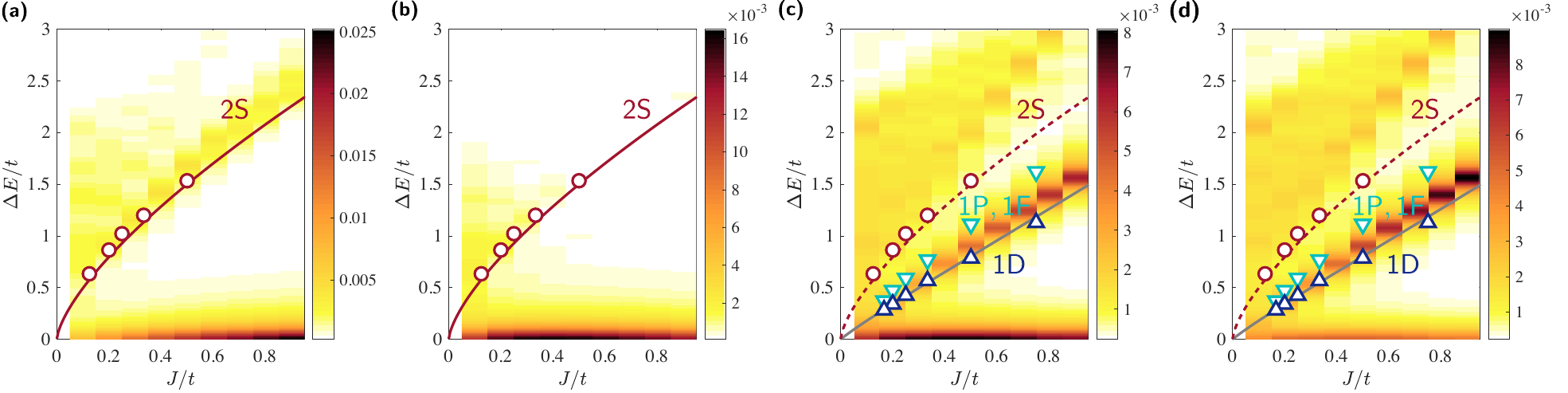, width=0.99\textwidth}
\caption{\textbf{Regge spectra from ED.} Using exact numerical diagonalization (ED) in a periodic $4 \times 4$ system, we calculate the rotational ARPES spectrum at the nodal point. We tune $J/t$ and plot energy distribution curves (color-coded) relative to the ground state energy at the respective value of $J/t$. We compare our numerical results with the DMRG and strong-coupling data from Fig.~\ref{figReggeTraj} in the main text (solid lines and symbols). The standard ARPES spectrum is shown in (a), rotational ARPES spectra for $m_4=0,1,2$ in (b)-(d). The same units (a.u.) were used for the different plots in (a)-(d).
}
\label{figReggeSpectraED}
\end{figure*}
\begin{figure*}[t!]
\centering
\epsfig{file=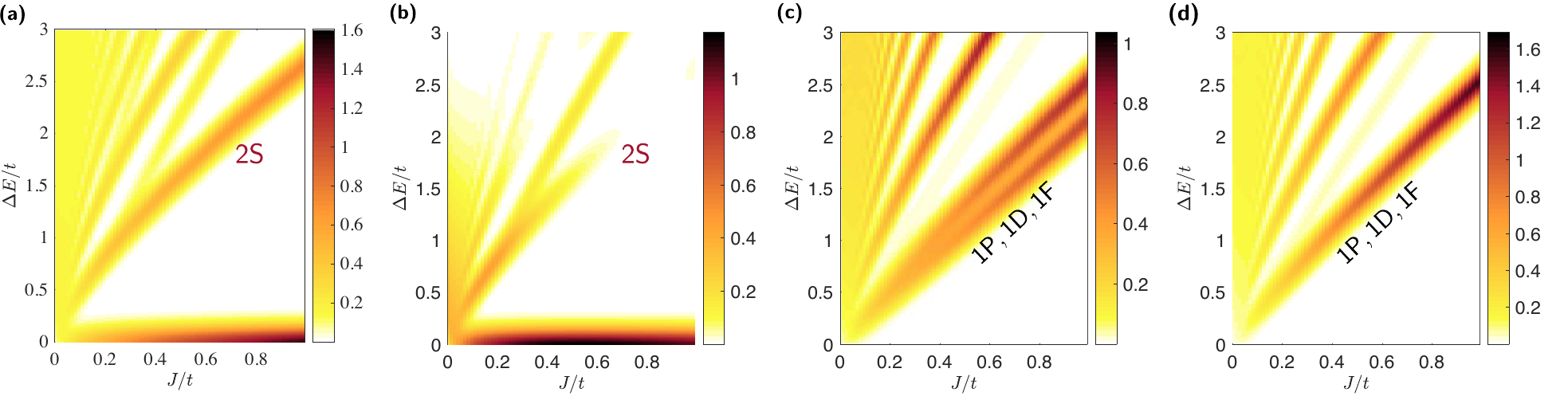, width=0.99\textwidth}
\caption{\textbf{Regge spectra from spinon-chargon toy model.} Using the spinon-chargon toy model, we calculate the rotational ARPES spectrum at the nodal point. As in Fig.~\ref{figReggeSpectraED} we tune $J/t$ and plot energy distribution curves (color-coded) relative to the ground state energy at the respective value of $J/t$. The standard ARPES spectrum is shown in (a), rotational ARPES spectra for $m_4=0,1,2$ in (b)-(d). The same units (a.u.) were used for the different plots in (a)-(d).
}
\label{figReggeSpectraToy}
\end{figure*}

\emph{Exact diagonalization (ED).--}
We have also performed numerically exact calculations of the (rotational) ARPES spectra in periodic $4 \times 4$ systems. Our results are in good agreement with our DMRG results presented in the main text. In particular, the rotational meson resonances are clearly visible in exact numerical studies even in the relatively small considered $4 \times 4$ systems.

We show our ED results in Fig.~\ref{figReggeSpectraED}, where we calculate the full spectrum at the nodal point $\vec{k}=(\pi/2,\pi/2)$ for different values of $J/t$. Note that all energies are measured relative to the respective ground state energy. The standard ARPES spectrum, Fig.~\ref{figReggeSpectraED} (a), and the trivial rotational spectrum at $m_4=0$, Fig.~\ref{figReggeSpectraED} (b), only show the lowest-lying vibrational resonance ($\mathsf{2S}$). For non-trivial $m_4=1,2$ the predicted rotational resonances are clearly visible in Fig.~\ref{figReggeSpectraED} (c) and (d). Furthermore, their excitation energy shows a clearly linear dependence on $J/t$ -- in accordance with the predicted Regge-trajectory from our toy model. 

We also compare the ED results in Fig.~\ref{figReggeSpectraED} with Regge trajectories extracted from our full DMRG calculations. Up to a few percent, the gap to the first vibrational excitation is identical in ED and DMRG. Both are in good agreement with the parameter-free strong coupling prediction, solid red line in Figs.~\ref{figReggeSpectraED} (a) and (b). The positions of the non-trivial rotational resonances at $m_4 \neq 1,2$ coincide in ED. This is different from the DMRG results for extended cylinders, where a pronounced splitting of $\mathsf{1D}$ and $\mathsf{1P}$ lines was found. We speculate that this is due to the special hypercubic symmetry special to the periodic $4 \times 4$ system, and thus not representative of larger systems. The overall position of the rotational peak agrees well with the DMRG results for extended cylinders, and with the parameter-free prediction by our strong coupling theory, solid gray line in Fig.~\ref{figReggeSpectraED} (c) and (d).  

In Fig.~\ref{figReggeSpectraToy} we show the same full Regge spectra at the nodal point, obtained from the spinon-chargon toy model. The results are in good qualitative agreement with our numerical data in Fig.~\ref{figReggeSpectraED}. The traditional APRES response in Fig.~\ref{figReggeSpectraToy} (a) closely matches our numerical results, including the rise of spectral weight of the $\mathsf{2S}$ resonance with increasing $J/t$. The trivial rotational APRES response ($m_4=0$) in Fig.~\ref{figReggeSpectraToy} (b) also agrees well with exact numerics at low energies. In particular it shows a strong suppression of spectral weight at the $\mathsf{2S}$ resonance when $J \geq 0.5 t$. This behavior is easily understood from the string picture: The $\mathsf{2S}$ vibrational excitation has one node in its radial string wavefunction. The position of the latter depends strongly on $J/t$. Since the rotational ARPES spectrum probes the overlap with string states of length $\ell=1$, the spectral weight vanishes when the node in the radial string wavefunction is exactly at $\ell=1$. Indeed, going to larger values of $J/t > 1$ (not shown), we find both for the toy model and in ED that spectral weight of the $\mathsf{2S}$ resonance re-appears; this is expected since the node in the radial string wavefunction moves even closer towards $\ell=0$.  

The toy model also correctly explains the qualitative shape of rotational Regge spectra at $m_4 \neq 0$, see Figs.~\ref{figReggeSpectraED} (c) and (d) and \ref{figReggeSpectraToy} (c) and (d). In particular, a weak second rotational resonance can be observed in Figs.~\ref{figReggeSpectraED} (c) and (d) which could be related to the second pronounced peak predicted by the toy model.

\begin{figure}[t!]
\centering
\epsfig{file=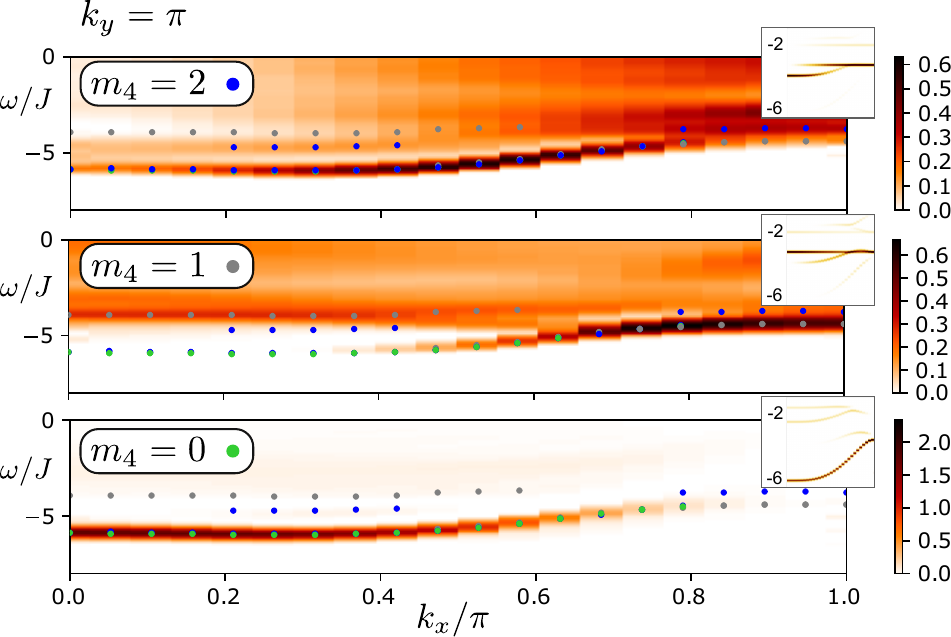, width=0.45 \textwidth}
\caption{\textbf{Rotational spectra and meson dispersion.} Using the same data as in Fig.~\ref{figSpectralLines} from the main text, we show the complete rotational ARPES spectrum, but for a cut at $k_y=\pi$. The insets show predictions by the spinon-chargon toy model for the same cut.
}
\label{figFullSpectraKyPi}
\end{figure}

\emph{Additional DMRG results.--}
From our time-dependent DMRG simulations, we have also extracted the full rotational ARPES spectra for straight cuts at $k_y=\pi$. Due to the broken translational symmetry of the undoped antiferromagnet, the eigenenergies at $(k_x,\pi)$ are expected to be identical to those at $(-\pi+k_x,0)$. These, in turn, are identical to those at $(\pi-k_x,0)$ due to inversion symmetry. Nevertheless, the spectral weights along the cut at $k_y=0$ in Fig.~\ref{figSpectralLines} from the main text, and along the cuts at $k_y=\pi$ shown here in Fig.~\ref{figFullSpectraKyPi} are different.

\subsection{Rotational meson states in 1D settings}
\label{SuppMat1Dmes}
Here we discuss a minimal toy model which illustrates rotational excitations and their signatures in ARPES spectra. We consider a situation where the charge dynamics is one-dimensional while the surrounding spins have a non-vanishing staggered magnetization. This case can be realized e.g. in 1D spin chains subject to a staggered magnetic field, or in mixed-dimensional situations \cite{Grusdt2018SciPost,Grusdt2020}, where signatures for mesonic spionon-chargon bound states have been predicted theoretically. The following treatment has direct experimental relevance in such cases.

\subsubsection{Model}
For concreteness we consider the following model Hamiltonian in a 1D doped spin chain,
\begin{equation}
 \H = J_z \sum_j \hat{S}^z_{j+1} \hat{S}^z_j +  h \sum_j (-1)^j \hat{S}^z_j  
 + \frac{J_\perp}{2} \sum_j \l \hat{S}^+_{j+1} \hat{S}^-_j + \hc  \r 
 - t ~ \hat{\mathcal{P}} \sum_{j,\sigma} \l  \cd_{j+1,\sigma} \c_{j,\sigma} +\hc \r \hat{\mathcal{P}}
 \label{eqToyModel}
\end{equation}
The first two terms describe AFM Ising couplings $J_z \geq 0$ between the spins in a staggered field $\pm h$, with $h>0$. The third term includes spin-exchange terms, and for simplicity we assume their amplitudes to be weak:
\begin{equation}
 J_\perp \ll J_z + h.
 \label{eq1SpnonCond}
\end{equation}
Finally, the last term describes hopping processes of doped holes, where $\hat{\mathcal{P}}$ is the projector to a subspace with $n=0,1$ fermions per lattice site. The operators $\c^{(\dagger)}_{j,\sigma}$ describe the underlying spin-$1/2$ fermions or bosons. Our discussion below is valid for arbitrary $t$, although we are most interested in the limit $t > J_z + h$. 

As the vacuum state we consider the half-filled N\'eel state without holes, $\ket{0} \equiv \ket{{\rm N}} = \ket{... \uparrow \downarrow \uparrow \downarrow ...}$, on top of which we will study one-hole excitations. When Eq.~\eqref{eq1SpnonCond} is satisfied, we can restrict ourselves to states with only one domain-wall, or spinon, excitation composed of aligned neighboring spins. The corresponding subspace is spanned by the following spinon-chargon states:
\begin{equation}
 	\hd_i \sd_j \ket{0} := \prod_{r=j}^i \left[ \sum_\mu \l \cd_{r,\mu} \c_{r+{\rm sgn}(i-j), \mu} \r \right] \c_{j,\sigma_j} \ket{\rm N},
\end{equation}
where $\sigma_j = \uparrow, \downarrow$ denotes the spin at site $j$ in the undoped N\'eel state. Here we introduced spinon and chargon operators $\hd_i$ and $\sd_j$ since we find working in second quantization more convenient. If the hole is left (right) from the domain wall, $i<j$ ($i>j$ respectively), the position $j$ of the spinon denotes the site of the left spin (right spin respectively) in the domain wall. When $i=j$, the hole is located in the middle between two aligned spins.

The effective Hamiltonian is obtained by projecting $\H$ from Eq.~\eqref{eqToyModel} to the subspace with one spinon and one chargon:
\begin{equation}
 \hat{\mathcal{P}}_{\rm 1s} \H \hat{\mathcal{P}}_{\rm 1s} = \H_{\rm sh}^0 + \H_{\rm sh}^{\rm int}. 
\end{equation}
The free spinon-chargon Hamiltonian is given by 
\begin{equation}
 \H_{\rm sh}^0 = - t \sum_i \hd_{i+1} \h_i  + \frac{J_\perp}{2} \sum_j  \sd_{j+2} \s_j  + \hc 
\end{equation}
and for the spinon-chargon interactions one finds:
\begin{equation}
\H_{\rm sh}^{\rm int} = \sum_i \sum_j \hd_i \h_i \sd_j \s_j V_{\rm sh}(|i-j|) 
- \frac{J_\perp}{2} \sum_j \hd_{j+1} \h_{j+1} \l \sd_{j+2} \s_j + \hc \r.
\label{eqHintSH}
\end{equation}
The first term describes the spinon-chargon potential
\begin{equation}
 	V_{\rm sh}(\ell) = h |\ell| - \delta_{\ell,0} \frac{J_z}{4},
\end{equation}
and the second term in Eq.~\eqref{eqHintSH} describes the absence of spinon motion when the hole is located in the middle of the domain wall formed by the spins. 

\subsubsection{Center-of-mass frame}
The spinon-chargon problem can be simplified by applying a unitary Lee-Low-Pines transformation \cite{Lee1953},
\begin{equation}
 	\hat{U}_{\rm LLP} = \exp \left[ - i \hat{X}_{\rm s} \hat{p}_{\rm h} \right]
\end{equation}
with the spinon position and chargon momentum
\begin{equation}
 \hat{X}_s = \sum_j j \sd_j \s_j, \quad \hat{p}_{\rm h} = \sum_k k \hd_k \h_k,
\end{equation}
which transforms into a reference frame co-moving with the spinon. The transformed Hamiltonian splits into blocks of fixed total momentum $K$:
\begin{equation}
	\hat{U}^\dagger_{\rm LLP} \H \hat{U}_{\rm LLP} = \bigoplus_{K} \tilde{\mathcal{H}}(K),
\end{equation}
where the individual blocks read:
\begin{equation}
  \tilde{\mathcal{H}}(K) = - t \sum_i \l \hd_{j+1} \h_j  + \hc \r + \sum_j V_{\rm sh}(|j|) \hd_j \h_j 
   + \frac{J_\perp}{2} \sum_{j \neq -1} \l e^{i 2 K} \hd_{j+2} \h_j + \hc  \r
   \label{eqHLLPtoy}
\end{equation}

\subsubsection{Symmetries}
The linear confining potential $V_{\rm sh}(\ell)$ binds the spinon to the chargon. In the strong coupling limit, defined by $J_\perp \ll t$, we can neglect the spinon motion and find that the resulting spinon-chargon Hamiltonian is inversion ($\hat{I}$) symmetric:
\begin{equation}
 	\hat{I}^\dagger  ~ \H |_{J_\perp = 0} ~ \hat{I} =  \H |_{J_\perp = 0},
\end{equation}
where the action of the inversion operator is $\hat{I}^\dagger \h_j \hat{I} = \h_{-j}$ and $\hat{I}^\dagger \s_j \hat{I} = \s_{-j}$. The inversion operator is the analogue of the discrete $C_4$ rotational symmetry in the 2D lattice. As a result, the spinon-chargon bound states $\ket{\psi_{\rm sh}^n}$ at strong coupling have a definite inversion quantum number: $\hat{I} \ket{\psi_{\rm sh}^n} = \xi_n \ket{\psi_{\rm sh}^n}$ with $\xi_n = \pm 1$.

When $J_\perp > 0$, away from the strong coupling limit, the spinon-chargon bound state is only inversion symmetric for certain inversion invariant momenta (IIM): $K = 0 , \pi/2, \pi$. Since $\hat{I}$ commutes with $\hat{U}_{\rm LLP}$, we can apply $\hat{I}$ to Eq.~\eqref{eqHLLPtoy} and obtain for general $K$:
\begin{equation}
	\hat{I}^\dagger ~ \tilde{\mathcal{H}}(K) ~ \hat{I} = \tilde{\mathcal{H}}( - K).
\end{equation}
I.e. spinon-chargon states $\ket{\tilde{\psi}_{\rm sh}^n(K)}$ have definite inversion eigenvalues $\xi_n(K) = \pm 1$ at IIM when $K_{\rm IIM} \equiv -K_{\rm IIM} {\rm mod} 2 \pi$: 
\begin{equation}
 \hat{I} ~ \ket{\tilde{\psi}_{\rm sh}^n(K_{\rm IIM})}~  = ~ \xi_n(K_{\rm IIM}) ~   \ket{\tilde{\psi}_{\rm sh}^n(K_{\rm IIM})} 
\end{equation}
Away from IIM, the spinon-chargon eigenstates will be superpositions of different inversion sectors in general.

\subsubsection{Spectral function}
Next we calculate the one-hole spectral function. In space and time coordinates it becomes,
\begin{equation}
 A_\sigma(j,t) = \bra{{\rm N}} e^{i \H t} \cd_{j,\sigma} e^{- i \H t} \c_{0,\sigma} \ket{{\rm N}}.
\end{equation}
Introducing spinon and chargon operators and LLP unitaries $\hat{U}_{\rm LLP}^\dagger \hat{U}_{\rm LLP} = 1$ yields:
\begin{equation}
 A(K,\omega) = {\rm Re} \frac{1}{\pi} \int_0^{\infty} dt ~ e^{i \omega t} \frac{1}{\sqrt{L}} \bra{0} \h_0 e^{-i  \tilde{\mathcal{H}}(K) t} \hd_0 \ket{0},
 \label{eqAkwgToy}
\end{equation}
where $L$ denotes the chain length.

\begin{figure}[t!]
\centering
\epsfig{file=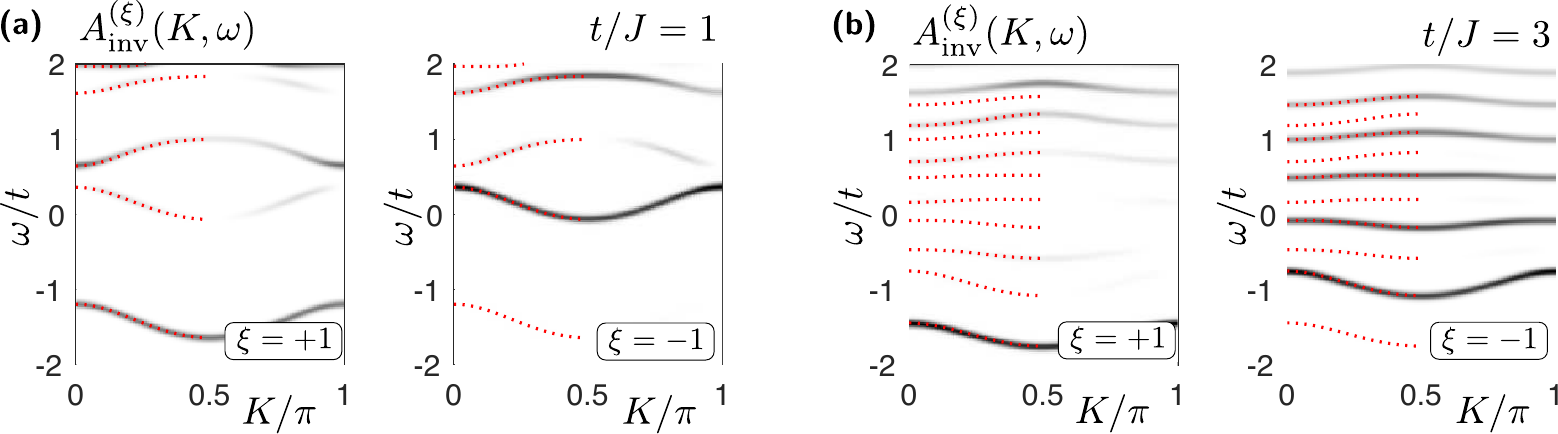, width=0.8 \textwidth}
\caption{\textbf{1D vibrational ARPES spectra} $A^{(\xi)}_{{\rm inv}}(K,\omega)$ are shown for the toy model described in the text. Parameters are $h=J_z=J_\perp=J$ with $t/J=1$ in (a) and $t/J=3$ in (b). At IIM $K=0,\pi/2,\pi$ the selection rules are exact. Dotted red lines indicate the positions of eigenstates and the spectra consist of delta-function peaks which have been slightly broadened for the illustration.  
}
\label{figRARPES}
\end{figure}

Because $\hd_0$ creates a chargon at the origin, this initial state in the time evolution of Eq.~\eqref{eqAkwgToy} has a definite positive inversion eigenvalue $\xi = +1$:
\begin{equation}
 	\hat{I} \hd_0 \ket{0} = (+1) ~ \hd_0 \ket{0}.
\end{equation}
As a result, the corresponding quasiparticle weight of odd-parity ($\xi=-1$) spinon-chargon bound states at IIM must vanish:
\begin{equation}
 	Z_n := |\bra{ \tilde{\psi}_{\rm sh}^n(K_{\rm IIM})} \hd_0 \ket{0}|^2 = 0 \quad \text{if} \quad \xi_n(K_{\rm IIM}) = -1.
\end{equation}
Such odd-parity states thus do not contribute to the spectrum at $K_{\rm IIM}$. For non-IIM the terms $\propto J_\perp$ in the Hamiltonian \eqref{eqHLLPtoy} explicitly break the inversion symmetry, and in general we expect non-zero spectral weight. The latter is small at strong couplings, however, when $J_\perp < t$ is weak. 

Likewise, we can calculate the analogue of the rotational spectrum, where the hole creation is immediately followed by a nearest-neighbor hopping with phases $1$ and $\xi$ along $\pm x$-direction:
\begin{equation}
 	A^{(\xi)}_{{\rm inv},\sigma}(j,t) =  \bra{{\rm N}} e^{i \H t} \cd_{j,\sigma} \hat{X}_\xi^\dagger(x) e^{- i \H t} \hat{X}_\xi(0) \c_{0,\sigma} \ket{{\rm N}},
\end{equation}
with $\xi = \pm 1$ and:
\begin{equation}
	\hat{X}_\xi(x) := \frac{1}{\sqrt{2}} \sum_{\mu = \uparrow, \downarrow} \cd_{x,\mu} \l \c_{x-1,\mu} + \xi \c_{x+1,\mu} \r.
\end{equation}
After changing into the Lee-Low-Pines frame as before and taking the Fourier transform to obtain the spectral function in momentum space, we obtain:
\begin{equation}
A^{(\xi)}_{{\rm inv},\sigma}(K,\omega) = {\rm Re} \frac{1}{\pi} \int_0^{\infty} dt ~ e^{i \omega t} \frac{1}{2 \sqrt{L}}  ~  \bra{0} \l \h_{-1} + \xi \h_1 \r e^{-i  \tilde{\mathcal{H}}(K) t} \l \hd_{-1} + \xi \hd_1 \r \ket{0}.
 \label{eqAkwgToyRot}
\end{equation}
In this case, the initial state in the time evolution of this equation has a definite parity $\xi = \pm 1$:
\begin{equation}
 \left[ \hat{I} -\xi \right] \frac{1}{\sqrt{2}} \l \hd_{-1} + \xi \hd_1 \r \ket{0} = 0.
\end{equation}

\end{document}